\documentclass[useAMS,usenatbib,fleqn]{mnras}
\usepackage{amsmath}
\usepackage{graphicx}
\usepackage{color}
\usepackage{natbib}
\usepackage{amssymb}

\bibpunct{(}{)}{;}{a}{}{,}

\def\rund#1{\left( #1 \right)}
\def\eps{{\epsilon}}
\def\vp{\varphi}

\def\ave#1{\left\langle #1 \right\rangle}
\def\vc#1{{\mbox{\boldmath$#1$\unboldmath}}}
\def\d{{\rm d}}

\title{Halo ellipticity of GAMA galaxy groups from KiDS weak lensing}

\author[Edo van Uitert et al.]
{Edo van Uitert$^{1}$\thanks{vuitert@ucl.ac.uk}, Henk Hoekstra$^2$, Benjamin Joachimi$^1$, Peter Schneider$^3$, \and Joss Bland-Hawthorn$^4$, Ami Choi$^5$, Thomas Erben$^3$, Catherine Heymans$^5$, \and Hendrik Hildebrandt$^3$, Andrew M. Hopkins$^6$, Dominik Klaes$^3$, Konrad Kuijken$^2$, \and  Reiko Nakajima$^3$, Nicola R. Napolitano$^7$, Tim Schrabback$^3$, Edwin Valentijn$^8$, \and Massimo Viola$^2$ \\ 
$^1$ Department of Physics and Astronomy, University College London, Gower Street, London WC1E 6BT, UK \\ 
$^2$ Leiden Observatory, Leiden University, Niels Bohrweg 2, NL-2333 CA Leiden, The Netherlands \\
$^3$ Argelander-Institut f\"ur Astronomie, Auf dem H\"ugel 71, 53121 Bonn, Germany \\ 
$^4$ Sydney Institute for Astronomy, School of Physics A28, University of Sydney, NSW 2006, Australia \\
$^5$ Scottish Universities Physics Alliance, Institute for Astronomy, University of Edinburgh, Royal Observatory, Blackford Hill, \\   Edinburgh EH9 3HJ, UK \\
$^6$ Australian Astronomical Observatory, PO Box 915, North Ryde, NSW 1670, Australia \\
$^7$ INAF - Observatory of Capodimonte, Salita Moiariello, 16, 800131, Naples, Italy \\
$^8$ Kapteyn Astronomical Institute, University of Groningen, PO Box 800, NL-9700 AV Groningen, the Netherlands \\
}

\pubyear{2016}

\begin{document}

\maketitle

\begin{abstract}
We constrain the average halo ellipticity of $\sim$$2\,600$ galaxy groups from the Galaxy And Mass Assembly (GAMA) survey, using the weak gravitational lensing signal measured from the overlapping Kilo Degree Survey (KiDS). To do so, we quantify the azimuthal dependence of the stacked lensing signal around seven different proxies for the orientation of the dark matter distribution, as it is a priori unknown which one traces the orientation best. On small scales, the major axis of the brightest group/cluster member (BCG) provides the best proxy, leading to a clear detection of an anisotropic signal. In order to relate that to a halo ellipticity, we have to adopt a model density profile. We derive new expressions for the quadrupole moments of the shear field given an elliptical model surface mass density profile. Modeling the signal with an elliptical Navarro-Frenk-White (NFW) profile on scales $R<$ 250 kpc, and assuming that the BCG is perfectly aligned with the dark matter, we find an average halo ellipticity of $\epsilon_{\rm h}=0.38\pm0.12$, in fair agreement with results from cold-dark-matter-only simulations. On larger scales, the lensing signal around the BCGs becomes isotropic and the distribution of group satellites provides a better proxy for the halo's orientation instead, leading to a 3--4$\sigma$ detection of a non-zero halo ellipticity at $250<R<750$ kpc. Our results suggest that the distribution of stars enclosed within a certain radius forms a good proxy for the orientation of the dark matter within that radius, which has also been observed in hydrodynamical simulations. 
\end{abstract}


\maketitle

%%%%%%%%%%%%%%%%%%%%%%%%%%%%%%%%%%%%%%%%%%%%%%%%%% Introduction %%%%%%%%%%%%%%%%%%%%%%%%%%%%%%%%%%%%%%%%%%%%%%%%%% 

\section{Introduction}
Simulations based on cold and warm dark matter models predict that dark matter collapses into anisotropic, approximately triaxial dark matter haloes, which appear elliptical in projection \citep[e.g.][]{Jing02,Maccio13}. Direct measurements of halo ellipticities therefore constitute a powerful test of the current cosmological paradigm, and serve as a test bed for modified gravity models, as some of these predict different halo ellipticity distributions \citep[e.g.][who studied the impact of a so-called `fifth force']{Hellwing13}, whilst others such as Modified Newtonian Dynamics \citep[MOND;][]{Milgrom83} generally predict a more axi-symmetric gravitational potential at larger distances from the galaxy where there are no baryons. Self-interacting dark matter models, on the other hand, lead to rounder haloes at small scales \citep{Peter13}. Therefore, halo ellipticity measurements as a function of scale have the potential to constrain the cross-section of dark matter. \\
\indent To constrain dark matter halo ellipticities, one could use visible tracers of the gravitational potential, such as the distribution and kinematics of satellite galaxies  \citep[e.g.][]{Brainerd05,Bailin08}, tidal streams of satellites \citep[e.g.][]{Ibata01,VeraCiro13}, H{\sc i} observations \citep[e.g.][]{Olling95,Peters16}, planetary nebulae \citep[e.g.][]{Hui95,Napolitano11}, strong lensing \citep[e.g.][]{Kochanek91,Caminha16} combined with stellar dynamics \citep[e.g.][]{VanDeVen10}, and X-ray observations \citep[e.g.][]{Fabricant84,Donahue16}. Most of these methods require assumptions on the relation between the dynamical state of the visible tracer and the dark matter in order to constrain halo ellipticities, which may lead to biases. In addition, tracers may not always be present in sufficient numbers at the scales of interest, for example towards the outer edges of the halo where dark matter dominates the total matter budget. These complications can be avoided with a different method: weak gravitational lensing. \\
\indent Weak lensing directly measures the total projected matter density around galaxies. A triaxial dark matter halo causes an azimuthal variation in the lensing signal; the amplitude is enhanced along the projected major axis, and decreased along the minor axis. This variation can be extracted from the lensing signal by applying a weight that varies with the angle between the image position vector, as measured from the lens centre, and the major axis of the matter distribution. \\
\indent The signal-to-noise ratio of weak lensing measurements for a single lens in typical wide-field surveys is far below unity for all but the most massive objects in the Universe. This is commonly dealt with by stacking the lensing signal around lenses of the same type. In order to coherently stack the anisotropic part of the lensing signal, we need to adopt a proxy for the major axis of the total matter distribution, as that is not known for individual lenses. Most studies to date used the observed major axis of the lens galaxy, relying on the assumption that galaxies and their haloes are aligned. In the extreme case that their orientations are uncorrelated, the anisotropy in the lensing signal is completely washed out. \\
\indent Most numerical simulations suggest a fairly large mean misalignment angle of $\vp_{\rm mis}\sim30^\circ$ between the galaxy and the dark matter halo, although the scatter between reported values is large \citep{VanDenBosch02,Okumura09,Bett10,Deason11,Bett12,Li13,Tenneti14}. Hydrodynamical simulations show that this misalignment angle depends on baryonic physics \citep{Tenneti16}, which make halo ellipticity constraints from weak lensing an interesting probe of baryonic feedback processes. A large misalignment would explain the reduction in the observed intrinsic alignment signal of luminous red galaxies compared to predictions from $\Lambda$CDM \citep{Faltenbacher09,Okumura09}, and could also explain why the weak-lensing-based halo ellipticity measurement of galaxy-scale haloes only reached tentative detections to date \citep{Mandelbaum06,VanUitert12,Schrabback15}. Recent hydrodynamical simulations, however, report that the misalignment angle decreases towards higher mass \citep{Tenneti15,Velliscig15}. In addition, N-body simulations have shown that more massive haloes are more elliptical on average \citep[e.g.][]{Allgood06,Despali14,Bonamigo15}. Indeed, for group-scale haloes and galaxy clusters, significant detections of anisotropy of the lensing signal have been reported \citep{Evans09,Oguri10,Clampitt16}. Hence these massive objects appear to be the optimal targets for lensing studies of halo ellipticity. \\
\indent This motivates us to measure the halo ellipticity of galaxy groups from the Galaxy And Mass Assembly (GAMA) survey \citep{Driver09,Driver11,Liske15}, using the imaging data from the Kilo Degree Survey \citep[KiDS;][]{DeJong13}. GAMA is a highly complete spectroscopic survey with $r<19.8$, which enabled the construction of a high-fidelity group catalogue \citep{Robotham11}. KiDS is an ongoing imaging survey optimized for weak lensing studies that includes complete coverage of GAMA. The combination of these two data sets is therefore perfectly suited for the task.\\
\indent Working with galaxy groups also has the advantage that we can measure the lensing signal anisotropy exclusively around group centrals only. Stacking the signal of central and satellite galaxies, as was presented in previous works on galaxy-scale haloes, also potentially dilutes the signal, and certainly complicates the interpretation. Another advantage of galaxy groups is that we can adopt various proxies for the orientation of the halo; not only can we use the major axis of the group central, but we can also use proxies based on the distribution of satellites, which also trace the dark matter \citep{Kang07,Wang08,Agustsson10,Dong14,Wang14}, or the vector connecting the group central to particular satellite galaxies. This enables us to investigate which proxy traces the dark matter orientation best.\\
\indent The outline of the paper is as follows. In Sect. \ref{sec_meth}, we outline the various estimators used to extract the average halo ellipticity from the lensing measurements. In Sect. \ref{sec_data} we discuss the data sets and outline the various proxies for the orientation of the dark matter distribution. The lensing measurements and the average halo ellipticities are presented in Sect. \ref{sec_res}. We conclude in Sect. \ref{sec_conc}. Throughout the paper we assume a standard $\Lambda$CDM cosmology with $\Omega_{\Lambda}=0.73$, $\Omega_{\rm M}=0.27$, and $h=0.7$ the dimensionless Hubble parameter, which is consistent with the best-fitting cosmological parameters from WMAP9 \citep{Hinshaw13}. We only explicitly write the Hubble parameter dependence of parameters in figures as a reminder. All distances are in physical (not co-moving) units.

%%%%%%%%%%%%%%%%%%%%%%%%%%%%%%%%%%%%%%%%%%%%%%%%%% Methodology %%%%%%%%%%%%%%%%%%%%%%%%%%%%%%%%%%%%%%%%%%%%%%%%%% 

\section{Methodology}\label{sec_meth}
\indent Weak gravitational lensing induces a small, coherent distortion in the shapes of background galaxies in the direction perpendicular to the separation vector between lens and source \citep[see][for a thorough introduction]{Bartelmann01}. This distortion is extracted by measuring the tangential projection of the observed ellipticities of background galaxies in concentric rings around the lens, the tangential shear:
\begin{equation}
\epsilon_{\rm t}= -\epsilon_1 \cos(2 \phi) - \epsilon_2 \sin (2 \phi),
\end{equation}
where $(\epsilon_1,\epsilon_2)$ are the two components of the observed galaxy ellipticity, which form an estimator of the gravitational shear $(\gamma_1,\gamma_2)$, and $\phi$ is the angle between the $x$-axis and the lens-source separation vector. The azimuthally averaged tangential shear is a useful quantity to measure as it is directly related to the differential mass profile of the lens:
\begin{equation}
  \langle\epsilon_{\rm t}\rangle(R) \approx \langle\gamma_{\rm t}\rangle(R) = \frac{\Delta\Sigma(R)}{\Sigma_{\mathrm{crit}}},
\end{equation}
with $R$ the projected separation between the lens and the source, $\Delta\Sigma(R)=\bar{\Sigma}(<R)-\langle \Sigma \rangle(R)$ the difference between the mean projected surface mass density inside $R$ and the azimuthally averaged surface mass density at $R$. $\Sigma_{\mathrm{crit}}$ is the critical surface mass density, which contains the geometric scaling of the lensing signal:
\begin{equation}
  \Sigma_{\mathrm{crit}}=\frac{c^2}{4\pi G}\frac{D_{\scriptscriptstyle\rm S}}{D_{\scriptscriptstyle\rm L} D_{\scriptscriptstyle\rm LS}},
\end{equation}
where $D_{\scriptscriptstyle\rm S}$, $D_{\scriptscriptstyle\rm L}$ and $D_{\scriptscriptstyle\rm LS}$ are the angular diameter distances to the source, to the lens, and between the source and the lens, respectively; $c$ is the speed of light and $G$ the gravitational constant.  \\
\indent The cross component of the shear is commonly measured as well:
\begin{equation}
  \epsilon_\times = \epsilon_1 \sin(2 \phi) - \epsilon_2 \cos (2 \phi).
\end{equation}
Weak gravitational lensing does not produce cross shear once azimuthally averaged. The cross shear is therefore commonly used as a null test to check for the presence of systematics. Elliptical mass profiles, however, can produce an azimuthally varying cross shear. \\ 

%%%%%%%%%%%%%%%%%%%%%%%%%%%%%%%%%%%%%%%%%%%%%%%%%% Methodology - anisotropic lens models %%%%%%%%%%%%%%%%%%%%%%%%%%%%%%%%%%%%%%%%%%%%%%%%%% 

\subsection{Anisotropic lens models}
Triaxial dark matter haloes cause an azimuthal variation in the lensing signal, which we want to extract. To relate that to a halo ellipticity, we have to adopt a lens model. We consider a mass distribution with confocal elliptical isodensity contours of axis-ratio $q\le 1$, 
\begin{equation}
\kappa(\vc\theta)=\ave{\kappa}\rund{\sqrt{{q \theta_1^2}+{\theta_2^2\over q}}}\;,
\label{eq_kappa}
\end{equation}
with $\kappa=\Sigma/\Sigma_{\rm crit}$ the convergence, where we choose coordinates such that the major axis lies along the $\theta_1$-direction. This parametrization ensures that the mass inside a isodensity contour $\kappa$ is independent of $q$. Defining the ellipticity of this distribution as \mbox{$\epsilon_{\rm h}=(1-q)/(1+q)$}, and expanding $\kappa$ in terms of $\epsilon_{\rm h}$ yields
\begin{eqnarray}
\kappa(\vc\theta)&=&\ave{\kappa}(\theta)-\epsilon_{\rm h}\,\theta\ave{\kappa}'\cos(2\vp)+{\cal O}(\epsilon_{\rm h}^2) \nonumber \\
&=:&\ave{\kappa}(\theta)+\epsilon_{\rm h}\,\kappa_2(\theta)\cos(2\vp)+{\cal O}(\eps_{\rm h}^2)\;,
\label{eq_kappaexp}
\end{eqnarray}
where in the final step we defined $\kappa_2(\theta)$; here, $\theta$ and $\vp$ are the polar coordinates of $\vc\theta$. In the following, we will neglect higher-order terms in $\epsilon_{\rm h}$; in this approximation, $\ave{\kappa}(\theta)$ is the azimuthally-averaged density profile of the lens. \\
\indent The deflection potential $\psi$ corresponding to the mass distribution of Eq. (\ref{eq_kappaexp}) reads \citep{Schneider91,Adhikari15}
\begin{equation}
\psi(\theta,\vp)=\psi_0(\theta)+\eps_{\rm h}\psi_2(\theta)\cos(2\vp)\;,
\label{eq_psi}
\end{equation}
with
\begin{equation}
\psi_0(\theta)=2\rund{\ln\theta\int_0^\theta\d\vartheta\;\vartheta\ave{\kappa}(\vartheta)+\int_\theta^\infty \d\vartheta\;\vartheta\ln\vartheta\ave{\kappa}(\vartheta)} \; ,
\label{eq_psi0}
\end{equation}
and
\begin{equation}
\psi_2(\theta)=-{1\over 2}\rund{{1\over \theta^2}\int_0^\theta\d\vartheta\;\vartheta^{3} \kappa_2(\vartheta) +\theta^2\int_\theta^\infty {\d\vartheta\over \vartheta}\kappa_2(\vartheta)}.
\label{eq_psi2}
\end{equation}
Since in the case considered here, $\kappa_2(\theta)=-\theta\,\ave{\kappa}'(\theta)$, the expression for $\psi_2$ can be simplified; using integration by parts, we obtain
\begin{equation}
\psi_2(\theta)=-{2\over\theta^2}\int_0^\theta\d\vartheta\;\vartheta^{3} \ave{\kappa}(\vartheta) \;.
\label{eq_psi2simp}
\end{equation}
This equation shows that the quadrupole term in the potential is affected only by the mass distribution inside the isodensity contour considered. In fact, it is known that the potential of lenses with elliptical isodensity contours depends only on the mass distribution inside an isodensity contour \citep{Bray84}, generalizing a corresponding result for axi-symmetric mass distributions. This property is preserved in all orders of the expansion of $\kappa$ (see Eq. \ref{eq_kappa}) in $\eps_{\rm h}$.\footnote{From Eq. (\ref{eq_kappaexp}) it is not obvious that this property also applies to the monopole term; however, as we shall see later, the derivatives of $\psi_0(\theta)$ only depend on $\ave{\kappa}(\theta')$ for $\theta'\le\theta$, and an additive constant of $\psi_0$ is arbitrary anyway.} \\
\indent One can check that this deflection potential satisfies the Poisson equation $\nabla_{\rm c}\nabla_{\rm c}^*\psi=2\kappa$, where we defined \mbox{$\nabla_{\rm c}={\partial/\partial\theta_1}+{\rm i}\, {\partial/\partial\theta_2}$}. Rewriting the differential operator in polar coordinates,
\begin{equation}
\nabla_{\rm c}={\rm e}^{{\rm i}\vp}\rund{{\partial\over\partial\theta}+{{\rm i}\over\theta}\,{\partial\over\partial\vp}}\;,
\label{eq_diffop}
\end{equation}
we find for the Laplacian of $\psi$
\begin{eqnarray}
\nabla_{\rm c}\nabla_{\rm c}^*\psi&=&{\partial^2\psi\over\partial\theta^2}+{1\over \theta}{\partial\psi\over\partial\theta}+{1\over\theta^2}{\partial^2\psi\over\partial\vp^2} 
\label{eq_lap1}
\\
&=&\psi_0''+{\psi_0'\over\theta}+\eps_{\rm h}\rund{\psi_2''+{\psi_2'\over\theta}-{4\psi_2\over\theta^2}}\;\cos(2\vp)\;.\nonumber
\label{eq_lap2}
\end{eqnarray}
To show that this equals $2\kappa$, we use 
\begin{equation}
{\psi_0'\over\theta}={2\over\theta^2}\int_0^\theta\d\vartheta\;\vartheta\,\ave{\kappa}(\vartheta)= \bar\kappa(\theta)\;,
\label{eq_kappabar}
\end{equation}
the mean surface mass density inside $\theta$, and \mbox{$\psi_0''=2\ave{\kappa}-\bar\kappa$}. Furthermore, \mbox{$\psi_2'(\theta)=-2\psi_2(\theta)/\theta -2\theta\ave{\kappa}(\theta)$} and \mbox{$\psi_2''(\theta)=6\psi_2(\theta)/\theta^2 +2\ave{\kappa}(\theta)-2\theta\ave{\kappa}'(\theta)$}.  With these relations, and using $\kappa_2=-\theta\ave{\kappa}'$, it is readily shown that $\nabla_{\rm c}\nabla_{\rm c}^*\psi=2\kappa$. \\
\indent The shear caused by the mass distribution of Eq. (\ref{eq_kappaexp}) is obtained in Cartesian coordinates as 
\begin{equation}
\gamma=\gamma_1+{\rm i}\,\gamma_2={1\over 2}\nabla_{\rm c}\nabla_{\rm c}\psi,
\label{eq_gamma}
\end{equation}
and its transformation to the tangential and cross components of the shear relative to the lens  center is defined as
\begin{equation}
\gamma_{\rm t}+{\rm i}\,\gamma_\times=-{\rm e}^{-2{\rm i}\vp}\gamma\;,
\label{eq_gammat}
\end{equation}
yielding\footnote{We note that the Eqs.\ (2.9) and (2.10) of \citet{Adhikari15} are missing a factor of 2.}
\begin{equation}
\gamma_{\rm t}={1\over 2}\rund{-{\partial^2\psi\over\partial\theta^2}+{1\over\theta}{\partial \psi\over\partial\theta}+{1\over\theta^2}{\partial^2\psi\over\partial\vp^2}} \; ; \; \gamma_\times={1\over\theta^2}{\partial\psi\over\partial\vp} -{1\over\theta}{\partial^2\psi\over\partial\theta\partial\vp}.
\label{eq_gammatop}
\end{equation}
Decomposing the shear into the monopole and quadrupole contributions,
\begin{eqnarray}
\gamma_{\rm t}(\theta,\vp)&=&\ave{\gamma_{\rm t}}(\theta) +\eps_{\rm h}\gamma_{\rm t,2}(\theta)\cos(2\vp)\;, \nonumber  \\
\gamma_\times(\theta,\vp)&=& \eps_{\rm h}\gamma_{\times,2}(\theta)\sin(2\vp) \;,
\label{eq_gammasep}
\end{eqnarray}
 -- note that the cross component has no monopole contribution -- we find that
\begin{eqnarray}
\label{eq_gamma_anis}
\ave{\gamma_{\rm t}}(\theta)&=&\bar\kappa(\theta)-\ave{\kappa}(\theta)\;,\nonumber\\
\gamma_{\rm t,2}(\theta)&=&-{6\psi_2(\theta)\over\theta^2}-2\ave{\kappa}(\theta)+\theta\ave{\kappa}'(\theta)\;,\\
\gamma_{\times,2}(\theta)&=&-{6\psi_2(\theta)\over\theta^2}-4\ave{\kappa}(\theta) \;.\nonumber
\end{eqnarray}
This is a simpler result than the one derived in \citet{Adhikari15}, as in order to calculate these shear functions, one only needs to solve a simple integral over a finite interval for $\psi_2$. \\
\indent The functions $\ave{\gamma_{\rm t}}(\theta)$, $\gamma_{\rm t,2}(\theta)$ and $\gamma_{\times,2}(\theta)$ are obtained from the shear components as
\begin{eqnarray}
\ave{\gamma_{\rm t}}(\theta)&=&{1\over 2\pi}\int_0^{2\pi}\d\vp\;\gamma_{\rm  t}(\theta,\vp) \;, \nonumber \\
\eps_{\rm h}\gamma_{\rm t,2}(\theta)&=&{1\over \pi}\int_0^{2\pi}\d\vp\;\gamma_{\rm t}(\theta,\vp)\,\cos(2\vp)\;,\\
\eps_{\rm h}\gamma_{\times,2}(\theta)&=&{1\over \pi}\int_0^{2\pi}\d\vp\;\gamma_{\times}(\theta,\vp)\,\sin(2\vp)\;.\nonumber
\end{eqnarray}
We now consider the estimators
\begin{eqnarray}
\widehat{\gamma}_{\rm t,0}(\theta)&=&{\sum_i w_i\,\eps_{\rm t,i} \over \sum_i w_i} \;,\nonumber \\
\widehat{\gamma}_{\rm t,2}(\theta)&=&{\sum_i w_i\,\eps_{\rm t,i}\, \cos(2\vp_i)\over \sum_i w_i\,\cos^2(2\vp_i)}\;, \\
\widehat{\gamma}_{\times,2}(\theta)&=&{\sum_i w_i\,\eps_{\times,i}\, \sin(2\vp_i)\over \sum_i w_i\,\sin^2(2\vp_i)} \nonumber,
\end{eqnarray}
where the sum extends over all image ellipticities within a given $\theta$ bin, and the $w_i$ are the shape measurement weights of the source galaxies. Furthermore, we drop the assumption that the major axis of the lens  is aligned with the $\theta_1$-axis, by interpreting the $\vp_i$ as the polar angle difference between a source galaxy's location and the lens major axis. Since the expection value of $\epsilon_i$ is ${\rm E}(\epsilon_i)=\gamma(\theta_i,\vp_i)$, we readily find from Eq. (\ref{eq_gammasep}), assuming that the source galaxy images have random polar angles $\vp_i$, that
\begin{eqnarray}
{\rm E}\rund{\widehat{\gamma}_{\rm t,0}(\theta)}&=&\ave{\gamma_{\rm t}(\theta)} \;,\nonumber \\
{\rm E}\rund{\widehat{\gamma}_{\rm t,2}(\theta)}&=&\eps_{\rm h}\,\gamma_{\rm t,2}(\theta)\; ,\\
{\rm E}\rund{\widehat{\gamma}_{\times,2}(\theta)}&=&\eps_{\rm h}\,\gamma_{\times,2}(\theta)\; . \nonumber
\end{eqnarray}
The convergence $\kappa(\vc\theta)$ and shear $\gamma(\vc\theta)$ of a lens depend of the distances of lens and source, due to the scaling of the physical surface mass density $\Sigma(\vc R)$ with the critical surface mass density $\Sigma_{\rm crit}$. Defining the distance-independent shear quantities $\Gamma_x(\vc R)=\Sigma_{\rm  crit}\,\gamma_x(D_{\scriptscriptstyle\rm L}\vc\theta)$, where $x$ denotes the various components of the shear, we can now combine the lensing signal of lenses and sources at different redshifts by defining the estimators
\begin{eqnarray}
\label{eq_Gammat0}
\widehat{\Gamma}_{\rm t,0}(R)&=&{\sum_i w_i\,\Sigma_{{\rm crit},i}^{-1}\,\eps_{{\rm t},i} \over \sum_i w_i\,\Sigma_{{\rm crit},i}^{-2}} \;,  \\
\label{eq_Gammat2}
\widehat{\Gamma}_{\rm t,2}(R)&=&{\sum_i w_i\,\Sigma_{{\rm crit},i}^{-1}\,\eps_{{\rm t},i}\,\cos(2\vp_i) \over \sum_i w_i\,\Sigma_{{\rm crit},i}^{-2}\,\cos^2(2\vp_i)} \;,\\
\label{eq_Gammax2}
\widehat{\Gamma}_{\times,2}(R)&=&{\sum_i w_i\,\Sigma_{{\rm crit},i}^{-1}\,\eps_{\times,i}\,\sin(2\vp_i) \over \sum_i w_i\,\Sigma_{{\rm crit},i}^{-2}\,\sin^2(2\vp_i)} \;,
\end{eqnarray}
where the sum extends over all source-lens pairs with transverse separation $R$ within a given bin. The weights are multiplied with a factor $\Sigma^{-2}_{\mathrm{crit}}$, which downweighs lens-source pairs that are close in redshift. These estimators have expectation values 
\begin{eqnarray}
{\rm E}\rund{\widehat{\Gamma}_{\rm t,0}(R) } &=& \Gamma_{\rm t,0}(R) = \Delta\Sigma(R) \;,\nonumber \\
{\rm E}\rund{\widehat{\Gamma}_{\rm t,2}(R) } &=&\eps_{\rm h}\, \Gamma_{\rm t,2}(R) \;,\\
{\rm E}\rund{\widehat{\Gamma}_{\times,2}(R) } &=&\eps_{\rm h}\, \Gamma_{\times,2}(R) \;. \nonumber
\end{eqnarray}
The $\Gamma_x(R)$ can be calculated for any given parametrized mass model $\Sigma(R)$, and these parameters, together with $\eps_{\rm h}$, can be determined by fitting these models to the estimators $\widehat{\Gamma}_x(R)$.\footnote{Comparing this to the lensing signal anisotropy model from \citet{Schrabback15},
\begin{equation}
  \gamma_{\rm{t}} (\vc\theta) = \langle \gamma_{\rm{t}} \rangle(\theta) [1+4 f_{\rm rel}(\theta) \epsilon_{\rm h} \cos(2\vp)] \;,\nonumber
\end{equation}
we identify \mbox{$4 \langle\gamma_{\rm t}\rangle (\theta) f_{\rm rel}(\theta)=-{6\psi_2(\theta)\theta^{-2}}-2\ave{\kappa}(\theta)+\theta\ave{\kappa}'(\theta)$}, while for the azimuthally varying cross term, we find $-4\langle\gamma_{\rm t}\rangle(\theta) f_{\rm rel,45}(\theta)=-{6\psi_2(\theta)\theta^{-2}}-4\ave{\kappa}(\theta)$. We have checked that $f_{\rm rel}(\theta)$ and $f_{\rm rel,45}(\theta)$, which are plotted in Fig. 2 of \citet{Mandelbaum06} for an elliptical Navarro-Frenk-White \citep[NFW;][]{Navarro96} profile and which were kindly provided to us in table format, satisfy these equations.}\\
\indent If the shape of the lens and the background sources are correlated, for example because of a spatially varying point spread function (PSF) pattern that is not fully corrected, or due to lensing by foreground structures (i.e. cosmic shear), $\epsilon_{\rm h}$ becomes biased if it is estimated from Eq. (\ref{eq_Gammat2}) only. Equation (\ref{eq_Gammax2}) is affected similarly by such a correlation but with an opposite sign, such that the sum of Eq. (\ref{eq_Gammat2}) and Eq. (\ref{eq_Gammax2}) is nearly unaffected \citep{Mandelbaum06,VanUitert12,Schrabback15}. Hence by fitting the average halo ellipticity to this sum, we can estimate the impact that lens-source alignments (either physical or systematic) have on our results. Our fiducial approach, however, is to fit these equations separately. We assess the difference in Sect. \ref{sec_sens}. \\
\indent To model the data and determine $\epsilon_{\rm h}$, we adopt an elliptical NFW profile to predict $\Delta\Sigma^{\rm NFW}$,  $\Gamma_{\rm t,2}^{\rm NFW}$ and $\Gamma_{\times,2}^{\rm NFW}$. 
We adopt a fixed mass-concentration relation from \citet{Duffy08}:
\begin{equation}
  c_{\mathrm{NFW}} = 5.71 \hspace{1mm} \Big( \frac{M_{200}}{2 \times 10^{12}h^{-1}M_{\odot}}\Big)^{-0.084} \hspace{1mm} (1+z)^{-0.47} \;,
  \label{eq_mass_c}
\end{equation}
with $M_{200}$ the mass inside a sphere with radius $r_{200}$, the radius where the density is 200 times the critical density $\rho_c$. Our results are not sensitive to the adopted mass-concentration relation, as shown in Sect. \ref{sec_sens}. Hence we are left with two fit parameters, the halo mass and the average halo ellipticity, $M_{200}$  and $\epsilon_{\rm h}$. We fit the isotropic (Eq. \ref{eq_Gammat0}) and anisotropic (Eq. \ref{eq_Gammat2} and Eq. \ref{eq_Gammax2}) part of the lensing signal simultaneously. We note that $\epsilon_{\rm h}$ can become negative, which implies that the dark matter distribution is oriented perpendicular to the major axis proxy. Also, as we have ignored terms of $\mathcal{O}(\epsilon^2_{\rm h})$ in Eq. (\ref{eq_kappaexp}), the $\epsilon_{\rm h}$ we obtain becomes increasingly biased towards larger ellipticities \citep[see Fig. 2 of][]{Schrabback15}.\\

%%%%%%%%%%%%%%%%%%%%%%%%%%%%%%%%%%%%%%%%%%%%%%%%%% Methodology -alternative estimator %%%%%%%%%%%%%%%%%%%%%%%%%%%%%%%%%%%%%%%%%%%%%%%%%% 

\subsection{Alternative estimator}\label{sec_alt}

We also implement the estimator of \citet{Clampitt16}, who propose to use the Cartesian components of the shear as the observables, which are defined in the reference frame where the $x$-axis is aligned with the major axis of the lens:
\begin{align}
\gamma_{1,2} & (\theta,\vp) = -\gamma_{\rm t,2}(\theta,\vp)\cos(2\vp) +\gamma_{\times,2}(\theta,\vp)\sin(2\vp) \;,\\
\gamma_{2,2} & (\theta,\vp) = -\gamma_{\rm t,2}(\theta,\vp)\sin(2\vp)-\gamma_{\times,2}(\theta,\vp)\cos(2\vp) \;. 
\end{align}
The advantage of these estimators compared to the ones defined in Eq. (\ref{eq_Gammat2}) and Eq. (\ref{eq_Gammax2}) is that a spurious, systematic alignment of lenses and sources only affects $\gamma_{1,2}$, but not $\gamma_{2,2}$, where it averages out \citep{Clampitt16}.\\
\indent These estimators are averaged in regions where $\cos(4\theta)$ (for $\gamma_{1,2}$) and $\sin(4\theta)$ (for $\gamma_{2,2}$) have the same sign:
\begin{eqnarray}
\label{eq_cl1}
\gamma^+_{1,2}(\theta) =: \frac{1}{\pi} \sum_{n=0}^3\int_{\pi/8+n\pi/2}^{3\pi/8+n\pi/2} {\rm d}\vp \;  \gamma_{1,2}(\theta,\vp) \;,\\
\gamma^-_{1,2}(\theta) =: \frac{1}{\pi} \sum_{n=0}^3\int_{-\pi/8+n\pi/2}^{\pi/8+n\pi/2} {\rm d}\vp \; \gamma_{1,2}(\theta,\vp) \;, \\
\gamma^+_{2,2}(\theta) =: \frac{1}{\pi} \sum_{n=0}^3\int_{\pi/4+n\pi/2}^{\pi/2+n\pi/2} {\rm d}\vp \;  \gamma_{2,2}(\theta,\vp) \;, \\
\label{eq_cl4}
\gamma^-_{2,2}(\theta) =: \frac{1}{\pi} \sum_{n=0}^3\int_{0+n\pi/2}^{\pi/4+n\pi/2} {\rm d}\vp \;  \gamma_{2,2}(\theta,\vp) \;.
\end{eqnarray}
 These integrals are estimated from the data as:
\begin{eqnarray}
\widehat{\Gamma}^{+/-}_{1,2}(R)=\frac{\sum_i w_i\Sigma_{{\rm crit},i}^{-1} [-\epsilon_{{\rm t},i}\cos(2\vp_i)+\epsilon_{\times,i}\sin(2\vp_i)]}{\sum_i w_i\Sigma_{{\rm crit},i}^{-2}}, \\
\widehat{\Gamma}^{+/-}_{2,2}(R)=\frac{\sum_i w_i\Sigma_{{\rm crit},i}^{-1} [-\epsilon_{{\rm t},i}\sin(2\vp_i)-\epsilon_{\times,i}\cos(2\vp_i)]}{\sum_i w_i\Sigma_{{\rm crit},i}^{-2}},
\end{eqnarray}
with the sum running over sources that fall inside a particular radial and azimuthal bin.\\
\indent We verified with mock elliptical NFW profiles that the halo ellipticity obtained with these estimators is the same as with our fiducial one (described in the previous section).

%%%%%%%%%%%%%%%%%%%%%%%%%%%%%%%%%%%%%%%%%%%%%%%%%% Data %%%%%%%%%%%%%%%%%%%%%%%%%%%%%%%%%%%%%%%%%%%%%%%%%% 

\section{Data}\label{sec_data}
This paper is one in a series that exploits the overlap between a highly complete spectroscopic survey, GAMA, and a deep optical imaging survey used for weak lensing, KiDS. Earlier work focused at the isotropic lensing signal to derive halo masses of galaxy groups, satellites in galaxy groups, galaxies as a function of stellar mass, and as a function of cosmic environment \citep[][respectively]{Viola15,Sifon15,VanUitert16,Brouwer16}. These works were based on 109 KiDS tiles that overlap with GAMA, from the first and second public data release to ESO \citep{DeJong15, Kuijken15}. In this work, we use the full overlap between KiDS and GAMA in the three equatorial patches (180 deg$^2$ in total). The shape and photometric redshift catalogue we use is a subset of the much larger KiDS-450 catalogue \citep{Hildebrandt16}. 

%%%%%%%%%%%%%%%%%%%%%%%%%%%%%%%%%%%%%%%%%%%%%%%%%% Data - GAMA %%%%%%%%%%%%%%%%%%%%%%%%%%%%%%%%%%%%%%%%%%%%%%%%%% 

\subsection{GAMA}\label{sec_gama}
GAMA is a highly complete spectroscopic survey that targeted galaxies with a Petrosian $r_{\rm AB}<19.8$ over roughly 286 deg$^2$ \citep{Driver09,Driver11,Liske15}. In this work, we only use data in three 12$\times$5 deg$^2$ patches near the equator, the so-called G09, G12 and G15 patches, for which a group catalogue has been made, G$^3$Cv7 \citep{Robotham11}. This group catalogue was constructed using a friends-of-friends algorithm that uses the projected and line-of-sight separation between galaxies, and has a high fidelity due to the completeness of GAMA \citep{Robotham11}. We use version 7 of that catalogue. We only use groups with five or more members, as a comparison with mock data has shown that an increasing fraction of groups are affected by interlopers towards lower group multiplicity \citep{Tankard15}.\\
\indent The group catalogue contains three different group centres: the brightest group/cluster galaxy (BCG), an iterative centre, which is obtained by computing the group centre of light and removing the galaxy furthest away from the centre iteratively, and the group centre of light. We adopt the BCG as the central galaxy as our fiducial set-up. \\
\indent We measure the shape of the BCG with the KSB method \citep{Kaiser95,LuppinoK97,Hoekstra98} on the KiDS data (described in Sect. \ref{sec_kids}), using the implementation described in \citet{Hoekstra98,Hoekstra00}. KSB determines the ellipticity of a galaxy using the higher-order moments of its brightness distribution. KSB is, as any shape measurement method, affected by noise bias \citep{Melchior12,Refregier12}, which depends on the brightness of the sample \citep{Hoekstra15}. Our BCGs are bright and large, however, and this bias should be small. Furthermore, the ellipticity estimates of the BCGs are mainly used to determine the major axis of the lens light, and the impact of noise bias on that is very small. Hence we do not apply a multiplicative bias correction to the ellipticities of the BCGs. As we discuss below, we measure the shapes of the source sample with a different shape measurement method, namely \emph{lens}fit \citep{Miller07,Kitching08}. Since the PSF is modeled using different prescriptions, any lens-source shape alignment due to inaccurate PSF modeling is less likely to be correlated, and hence suppressed. Furthermore, since the BCGs are very large and bright in the KiDS imaging data, the PSF has only a small impact on their shapes, which further reduces the chance of a spurious lens-source alignment. We visually inspect all BCGs with a shape measurement that fall inside a KiDS mask. Roughly a quarter of those are close to a bright star or affected by image artefacts and are excluded, but the others are kept in the analysis. \\
\indent In addition to the major axis of the BCG, we can use various other proxies for the orientation of the projected dark matter distribution that might be better aligned with it, based on the positions of the satellite galaxies in each group. We start with defining the quadrupole moments of the cluster member distribution:
\begin{equation}
Q_{ij} =\frac{\sum_k (\theta_{i,k}-\theta_{i}^{\rm BCG})(\theta_{j,k}-\theta_{j}^{\rm BCG}) w(R_k)}{\sum_k w(R_k)}, \hspace{2mm} i,j \in \{1,2\},
\label{eq_mom}
\end{equation}
where the sum runs over the satellites in the group, $(\theta_{1,k},\theta_{2,k})$ is the position of satellite $k$ in the image plane and $(\theta_1^{\rm BCG},\theta_2^{\rm BCG})$ is the position of the BCG in the image plane. $w(R_k)$ is a radial weight function that depends on the projected separation between the BCG and the satellite. We adopt this weight function to suppress noise caused by potential field galaxy interlopers at large distances from the BCG. We try two weight functions: a Gaussian with a projected scale radius of $R_w=300$ kpc (roughly equal to $0.5 \; r_{200}$) and a Gaussian with a projected scale radius of $R_w=600$ kpc, as it is not a priori clear how to best suppress the effect of possible interlopers. It potentially also allows us to explore whether the satellites near the group centre, or the ones near the edge, trace the dark matter distribution better. From these moments we form the complex ellipticity of the group member distribution:
\begin{equation}
\epsilon_{\rm sat}=\frac{Q_{11}-Q_{22}+2{\rm i}Q_{12}}{Q_{11}+Q_{22}+2\sqrt{Q_{11}Q_{22}-Q_{12}^2}} \;. \\
\label{eq_esat}
\end{equation}
We address the noise bias due to the use of a relatively small number of satellites below. Note that this estimator is similar to the one adopted in \citet{Evans09}, with the exception of the weight function. In the absence of interlopers, Eq. (\ref{eq_esat}) is biased low compared to the true ellipticity of the group member distribution, because of the use of a weight function. This weight function should therefore be included in any comparison with simulations. \\
\begin{figure}
   \centering
   \includegraphics[width=1.2\linewidth,angle=270]{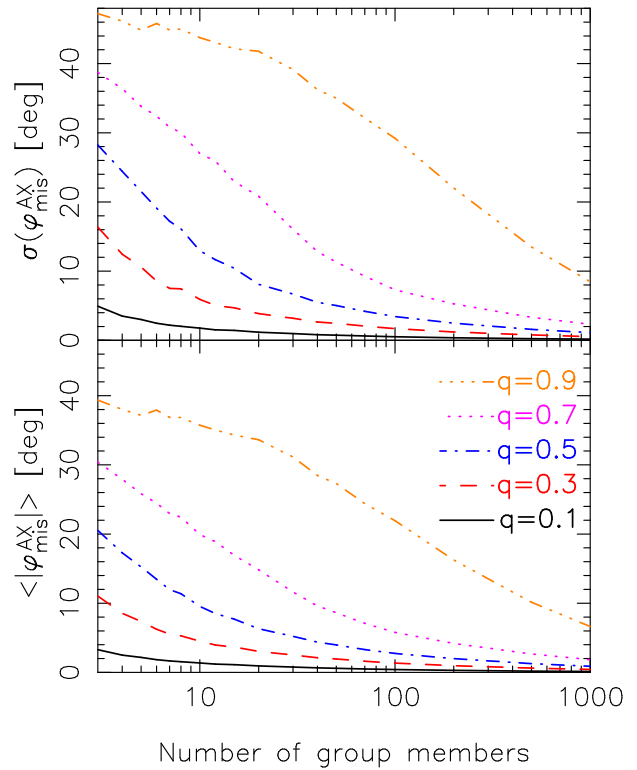}
   \caption{Standard deviation of the misalignment distribution (top) and mean of the absolute value of the misalignment angle (bottom) as a function of the number of satellite galaxies sampling the dark matter distribution. Various lines correspond to different axis ratios $q$ of the projected dark matter distribution. The misalignment distribution becomes wider when the haloes become rounder, and when there are fewer satellites to sample the matter distribution. The mean of the absolute value of the misalignment angle has an upper bound of $45^\circ$ (to within the errors) if the angles are uncorrelated, but the standard deviation can reach values larger than that (e.g. if the misalignment probability distribution is flat).}
   \label{plot_nfwsamp}
\end{figure}
\indent Red satellite galaxies are more clustered around the major axes of central galaxies than blue satellites \citep[e.g.][]{Yang06,Huang16}. The distribution of red satellites may therefore be a better tracer of the orientation of the projected dark matter distribution. Hence we also use the distribution of red group members exclusively as another proxy, selected using their rest-frame $(u-r)$ colours derived from the spectral energy distribution fits to the photometry to estimate stellar masses \citep{Taylor11}. We adopt $(u-r)>1.8$, which roughly corresponds to the bimodality scale in the colour-magnitude relation. Approximately 70\% of the satellites are red according to this criterion. We only use groups with three or more red members, which is the minimum number required for the moments in Eq. (\ref{eq_mom}) to be independently determined. $\sim$23\% of the groups have fewer than three red members and they are not used when we adopt this estimator. The proxies are listed in Table \ref{tab_he}. \\
\begin{figure}
   \centering
   \includegraphics[width=1\linewidth]{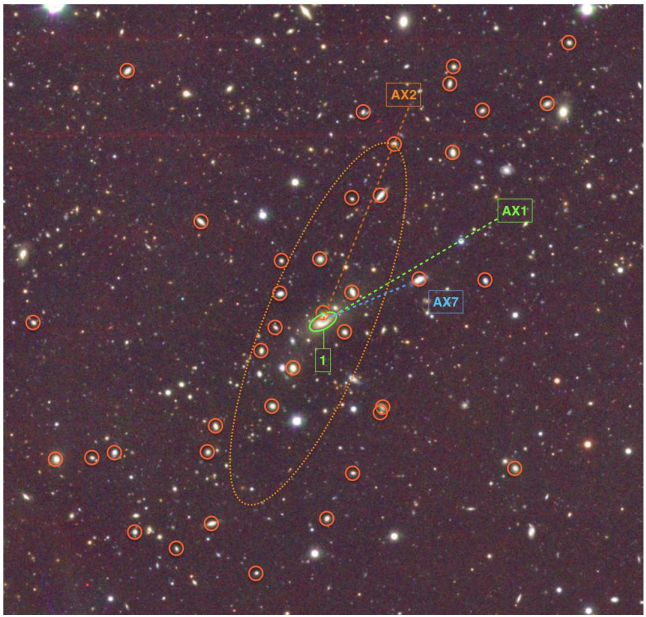}
   \caption{Illustration of the axes around which we measure the lensing signal anisotropy. Depicted is a false colour image based on KiDS $u,g,r,i$-band data of GAMA group 200004 at $\alpha,\delta=(182.3333,-2.2251)$. The green ellipse indicates the BCG, the orange circles the group members, and the large dotted ellipse the ellipticity of the group member distribution. The dashed lines indicate three of the seven axes around which we measure the lensing signal anisotropy: the green one is defined by the major axis of the BCG (AX1), the orange one is along the major axis of the group member distribution (AX2) and the blue one is defined by the separation vector of the BCG and the third brightest satellite (AX7). The brightest and second brightest satellite just fall outside the plotted range (towards the bottom left) and hence AX5 and AX6 are not shown. AX3 and AX4 are also not shown as they are very similar to AX2. The image roughly measures $10\times10$ arcmin on a side, which corresponds to 1.8 Mpc at the BCG's redshift of 0.176.}
   \label{plot_ill}
\end{figure}
\begin{figure}
   \includegraphics[angle=270,width=1\linewidth]{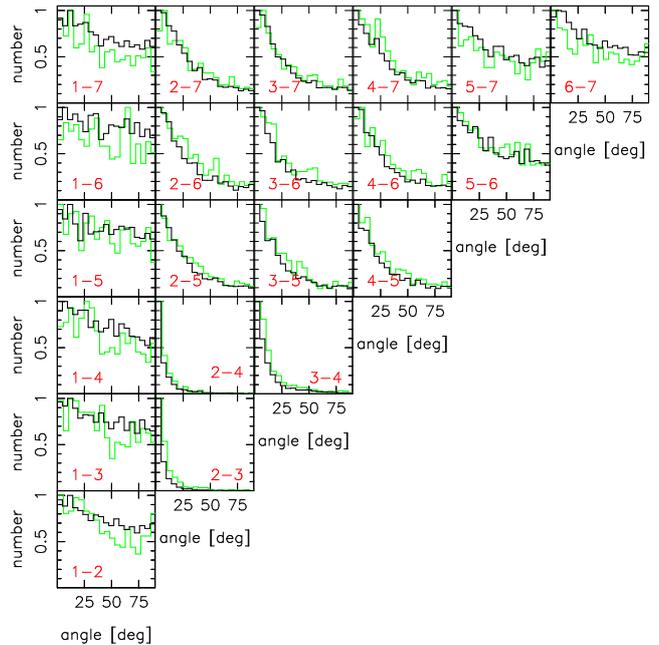}
   \caption{Distribution of the absolute values of the angles between the axes around which we measure the lensing signal anisotropy, as defined in Table \ref{tab_he}. The horizontal axis of each panel shows the angle between the two axes in degrees, the vertical axis the number of times this angle is observed in GAMA groups with $N_{\rm fof}\ge5$ (normalised to unity). The red numbers in each panel correspond to the axes names (see Table \ref{tab_he}). Black lines are for all groups with $N_{\rm fof}\ge5$, green lines are for groups with $N_{\rm fof}\ge10$.}
   \label{plot_prefax}
\end{figure}
\indent The low number of satellites per group makes the satellite distribution a noisy proxy of the orientation of the dark matter distribution, even if satellites perfectly trace the dark matter. To estimate the misalignment distribution purely due to sampling variance, we sample from a 3D NFW profile with a range of axis ratios, where the sampling probability is proportional to the density. The BCG is placed at the centre of the halo. We compare the axis ratio of the projected dark matter distribution to the one estimated from the sampled satellites. The axis ratios are determined without a radial weight function, which has virtually no effect as there are no interlopers in these simulations. The result is shown in Fig. \ref{plot_nfwsamp}: the distribution of satellites is a noisy proxy for the halo's major axis when the halo is round, and when the number of satellites is low. We use the observed distribution of group multiplicities and axis ratios to determine the mean standard deviation of the misalignment distribution, which we need in order to estimate how much the halo ellipticity estimates are diluted by this sampling variance. For AX2 and AX3 (the major axis of the satellite distribution determined using a weight function with a scale radius of 300 kpc and 600 kpc, respectively; see Table \ref{tab_he}), the minimum number of satellites that we use is four, while for AX4 (the major axis of the distribution of red group members) the minimum number is three. The mean standard deviations $\sigma(\vp_{\rm mis}^{\rm AX})$ are 21.0, 18.5, and 23.0 degrees for AX2, AX3 and AX4, respectively, with an estimated error of 1.0 degree due to the random sampling of the haloes (i.e. due to simulating a finite number of haloes). \\
\indent If we stack the lensing signal around an axis that differs by $\vp_{\rm mis}$ from the orientation of the dark matter distribution, the anisotropic part of the lensing signal becomes proportional to $\cos(2[\vp + \vp_{\rm mis}])=\cos(2\vp)\cos(2\vp_{\rm mis})-\sin(2\vp)\sin(2\vp_{\rm mis})$. $\sin(2\vp)$ averages to zero, hence the second term vanishes. The lensing signal anisotropy is therefore diluted by a factor $\cos(2\vp_{\rm mis})$. If the misalignment distribution is Gaussian, the dilution of the average halo ellipticity is given by:
\begin{equation}
D=1-\frac{\int_{0}^{\pi} {\rm d}\vp_{\rm mis} \exp{(-0.5[\vp_{\rm mis}/\sigma(\vp_{\rm mis}^{\rm AX})]^2)} \cos(2\vp_{\rm mis})}{\int_{0}^{\pi} {\rm d}\vp_{\rm mis} \exp{(-0.5[\vp_{\rm mis}/\sigma(\vp_{\rm mis}^{\rm AX})]^2)}}\;,
  \label{eq_dil}
\end{equation}
which has a value of 24\%, 19\% and 28\% for AX2 to AX4, respectively. These numbers are only indicative, as we did not account for the relative lensing weight of groups with different masses, the misalignment distribution may not be Gaussian, and the assumption that satellites perfectly trace the dark matter may be incorrect. Hence we do not correct our results with these factors, but we note that they should be kept in mind when comparing results. \\
\indent The lower panel of Fig. \ref{plot_nfwsamp} shows the mean absolute misalignment angle, which is a quantity that is frequently reported in studies of the satellite distribution of galaxy groups and clusters. The mean absolute misalignment angle and the standard deviation of the misalignment distribution are complementary as the relation between the two depends on the misalignment distribution itself.\\
\indent The final three proxies for the orientation of the projected dark matter distribution are the separation vectors from the BCG to the three brightest satellite galaxies. The magnitude we use to rank the satellites is the $r_{\rm AB}$ Petrosian apparent magnitude from SDSS. \\
\indent In total, we use seven different proxies for the orientation of the dark matter distribution. We illustrate the difference between some of the proxies in Fig. \ref{plot_ill} and list all of them in Table \ref{tab_he}. The total number of lenses used for the AX1 and AX4 samples is slightly lower than the rest. For AX1, the reason is that not all BCGs have reliable shape measurements. For AX4, some groups have fewer than three red satellites and are therefore excluded. \\
\begin{table*}
  \caption{Halo ellipticity constraints around different proxies for the orientation of the dark matter distribution. The first column contains the name of the proxy, the second column a brief description, the third column the number of lenses used, and the fourth to sixth column the constraints on the halo ellipticity, obtained by fitting an elliptical NFW profile over different radial ranges.}   
  \centering
  \begin{tabular}{c c c c c c} 
  \hline
Axis name &  Preferred axis & ${\rm N_{lens}}$ &  $\epsilon_{\rm h}$ & $\epsilon_{\rm h}$ & $\epsilon_{\rm h}$ \\
 &  &  &  40 kpc $<R<$ 250 kpc & 40 kpc $<R<$ 750 kpc & 250 kpc $<R<$ 750 kpc \\
  \hline\hline
\\
 AX1 & BCG orientation & 2355 & $0.38\pm0.12$ & $0.24\pm0.08$ & $0.05\pm0.13$\\
 AX2 & sat. dstr. ($R_w$=300 kpc) & 2672 &  $-0.04\pm0.11$ & $0.17\pm0.08$ & $0.49\pm0.13$ \\
 AX3 & sat. dstr. ($R_w$=600 kpc) & 2672 &  $0.05\pm0.11$ & $0.19\pm0.08$ &  $0.42_{-0.12}^{+0.13}$\\
 AX4 & sat. dstr. ($R_w$=300 kpc, red) & 2051 &  $0.03\pm0.11$ & $0.18_{-0.07}^{+0.08}$ & $0.39_{-0.11}^{+0.12}$\\
 AX5 & vector BCG - brightest sat. & 2672 &  $0.25\pm0.11$ & $0.25\pm0.08$ &  $0.27_{-0.12}^{+0.13}$\\
 AX6 & vector BCG - 2$^{\rm nd}$ brightest sat & 2672 &  $-0.09\pm0.11$ & $-0.10\pm0.08$ &$-0.11\pm0.12$\\
 AX7 & vector BCG - 3$^{\rm rd}$ brightest sat & 2672 &  $-0.12\pm0.11$ & $-0.05\pm0.08$ & $0.06\pm0.12$\\
\\
  \hline
  \end{tabular}
  \label{tab_he}
\end{table*} 
\begin{table*}
  \centering
  \caption{Mean absolute angle between the axes of the different proxies around which we measure the lensing signal anisotropy. The first set of rows show the results for groups with $N_{\rm fof}\ge5$, the second set is for groups with $N_{\rm fof}\ge10$.}
  \renewcommand{\tabcolsep}{0.6cm}
  \begin{tabular}{c c c c c c c} 
  \hline
Axis name &  AX2 & AX3 & AX4 & AX5 & AX6 & AX7 \\
  \hline\hline
\\
 $N_{\rm fof}\ge5$ &  &  &  &  &  &  \\
\\
 AX1 & $41.7\pm0.5$ & $42.0\pm0.5$ & $40.8\pm0.6$ & $42.6\pm0.5$ & $43.1\pm0.5$  & $41.3\pm0.5$\\
 AX2 & - & $7.4\pm0.2$  & $9.2\pm0.3$  & $26.3\pm0.4$  & $27.9\pm0.5$  & $28.0\pm0.4$  \\
 AX3 & - & - & $13.8\pm0.4$ & $26.0\pm0.5$ & $27.8\pm0.5$ & $28.3\pm0.5$  \\
 AX4 & - & - & - & $26.4\pm0.5$ & $28.4\pm0.5$ & $29.7\pm0.5$  \\
 AX5 & - & - & - & - & $37.8\pm0.5$  & $38.3\pm0.5$  \\
 AX6 & - & - & - & - & - & $39.8\pm0.5$ \\
\\
\hline
\\
 $N_{\rm fof}\ge10$ &  &  &  &  &  &  \\
\\
 AX1 & $39.9\pm1.2$ & $40.4\pm1.1$ & $39.5\pm1.1$ & $41.4\pm1.1$ & $42.2\pm1.1$ & $39.2\pm1.1$ \\
 AX2 & - & $10.8\pm0.6$ & $10.1\pm0.6$ & $28.3\pm1.0$ & $30.6\pm1.0$ & $30.9\pm1.0$ \\
 AX3 & - & - & $15.7\pm0.7$ & $27.1\pm1.0$ & $29.8\pm1.0$ & $30.8\pm1.0$ \\
 AX4 & - & - & - & $28.0\pm1.0$ & $31.5\pm1.0$ & $31.5\pm1.0$ \\
 AX5 & - & - & - & - & $38.0\pm1.1$ & $40.0\pm1.1$ \\
 AX6 & - & - & - & - & - & $40.3\pm1.1$ \\
  \hline
  \end{tabular}
  \label{tab_dtheta}
\end{table*} 
\indent We show the distribution of relative angles between the various major axis proxies in Fig. \ref{plot_prefax}. The first column shows the distribution of angles between the major axis of the BCG and the various proxies based on the distribution of satellites. It shows that satellites preferentially reside near the major axis of the BCG. We report the mean absolute angle (not the standard deviation of the misalignment distribution) in Table \ref{tab_dtheta}. The quoted errors on the average angle are the standard error of the mean. The mean absolute angle between the major axis of the BCG and the one of the distribution of satellites, estimated using Eq. (\ref{eq_esat}), is $\sim$$41.7\pm0.5$ degrees, and is insensitive to the scale radius of the weight function. This agrees well with previous results reported in the literature \citep[see e.g.][]{Yang06,Wang08,Huang16}.  \\
\indent This result, together with Fig. \ref{plot_nfwsamp} and the corresponding mean absolute misalignment angles, shows that the major axis of the BCG and the major axis of the satellite distribution cannot both trace the orientation of the dark matter distribution well, as the average absolute angle is much larger than the contribution from sampling variance. Our measurements of the lensing signal anisotropy will reveal which one is the better tracer of the orientation of the halo. \\
\indent The other panels of Fig. \ref{plot_prefax} and Table \ref{tab_dtheta} also contain a wealth of information about the satellite distribution. We will explore this in a future study, as this is outside the scope of this work.

%%%%%%%%%%%%%%%%%%%%%%%%%%%%%%%%%%%%%%%%%%%%%%%%%% Data - KiDS %%%%%%%%%%%%%%%%%%%%%%%%%%%%%%%%%%%%%%%%%%%%%%%%%% 

\subsection{KiDS}\label{sec_kids}
KiDS is a currently ongoing, large, optical imaging survey specifically designed for weak lensing studies. The survey will eventually map 1500 deg$^2$ of sky in four optical bands ($u$, $g$, $r$, and $i$) to a depth of 24.3, 25.1, 24.9, 23.8 (5$\sigma$ in a 2 arcsec aperture), respectively. Additional coverage in five infrared bands from the VISTA Kilo-degree Infrared Galaxy (VIKING) survey \citep{Edge13} will enable exquisite photometric redshifts estimates as well as stellar mass estimates for millions of galaxies.  The lensing measurements are performed on the $r$-band exposures, which are taken under the most stringent seeing condition ($<0.8''$). \\
\indent In this work, we used the KiDS-450 catalogue \citep{Hildebrandt16}, which superseded the KiDS-DR1/2 catalogues \citep{DeJong15,Kuijken15}. As detailed in \citet{Hildebrandt16}, the KiDS-450 catalogues have improved in various ways compared to the KiDS-DR1/2 catalogues. An updated, `self-calibrating' version of the shape measurement method \emph{lens}fit \citep{Miller07,Kitching08} has been employed \citep{Fenech16} to measure the shapes of the source galaxies, which has an average multiplicative bias of only $\sim$1\% for galaxies in the range 0.1$<$$z_B$$\le$0.9, with $z_B$ the Bayesian point estimate of the photo-$z$. Furthermore, the KiDS-450 catalogues do not contain the redshift probability distribution of individual galaxies as estimated from BPZ \citep{Benitez00,Hildebrandt12}, as those were found to be biased. Instead, the total source $N(z)$ is estimated through a weighted direct calibration technique based on the overlap with deep spectroscopic samples \citep{Hildebrandt16}. We determine the $N(z)$ for every lens redshift separately, by selecting all galaxies in the spectroscopic sample with a $z_B$ larger than $z_{\rm lens}+0.1$ and determining their spectroscopic redshift distribution. The same source redshift cut is applied in the lensing analysis (see below). To determine $\Sigma_{\rm crit}$ for a given lens redshift, we integrate over this $N(z)$. Since we multiply the \emph{lens}fit weights with a factor $\Sigma^{-2}_{\rm crit}$ in Eq. (\ref{eq_Gammat0})--(\ref{eq_Gammax2}), we effectively upweigh groups at lower redshift. The KIDS-450 catalogues have a source density of $n_{\rm eff}=8.53$ galaxies/arcmin$^2$, significantly higher than in the KiDS-DR1/2 catalogues due to the implementation of improved galaxy deblending criteria. \\
\indent The quality of the lensing catalogues has been thoroughly assessed in \citet{Hildebrandt16} and was found to be sufficient for cosmic shear measurements. A number of additional, galaxy-galaxy lensing specific tests have been performed (Dvornik et al., in prep.). As shown there, the lensing signal around random points is consistent with zero on the scales of interest in this work ($<$1 Mpc). Hence we do not subtract the signal around random points to correct for systematics. Secondly, the KiDS-450 catalogues contain a small but non-negligible additive bias. If both the lens and source galaxy are affected by a similar additive bias, the lensing anisotropy measurements become biased, which can be corrected for by fitting to the sum of Eq. (\ref{eq_Gammat2}) and (\ref{eq_Gammax2}). However, since the shapes of our lenses are measured with an independent pipeline that uses a different PSF model, it is unlikely that the additive biases are strongly correlated. Finally, \citet{Fenech16} report a remaining multiplicative bias of the order of a percent in the shape measurement catalogues. Since this bias does not affect the relative scaling of the isotropic and anisotropic part of the lensing signal (which constrains the halo ellipticity), nor does it affect the azimuthal dependence of the lensing signal, we do not correct for it. \\
\indent To remove source galaxies that are physically associated to the galaxy groups, we apply a redshift cut based on $z_B$. As mentioned before, we require $z_B$(source)$>z_{\rm lens}+0.1$. To maximize the source number density, we do not apply an upper limit to the source redshifts of $z_B\le0.9$, as recommended in \citet{Hildebrandt16}. As a result, our $N(z)$ may be somewhat biased, and in addition, the average multiplicative bias may be somewhat larger than $\sim$0.01, but since this affects the isotropic and anisotropic part of the lensing signal equally, our constraints on halo ellipticity should not be biased as they effectively depend on the ratio of the two. The halo masses we report might be biased, but they only serve as a nuisance parameter in the fit. Reliable estimates of the halo masses of these groups can be found in \citet{Viola15}. A small, secondary effect is that the scale radius of the NFW profile that we use in the fit might be slightly biased. To test if that affects the results, we fit the data using a different normalization of the mass-concentration relation and find this has no significant effect (see Sect. \ref{sec_sens}).\\
\indent The photometric redshift probability distribution of individual source galaxies is broad, and a significant fraction of sources will be physically associated with the lenses even after applying $z_B$(source)$>z_{\rm lens}+0.1$. This biases the shear measurements as a function of projected radius, which would bias our results if unaccounted for. To correct for this, we determine the overdensity of source galaxies around the lens, and multiply this `boost factor' with the shear measurements \citep[e.g.][]{Mandelbaum06,VanUitert11}. We use the same lens-source pair weights when we compute the overdensity, that is the product of the \emph{lens}fit weight of the source galaxy and $\Sigma_{\rm crit}^{-2}$. The boost factor reaches a maximum of $\sim$1.1 around 75 kpc, and decreases with projected radius to values below 1.05 at $\sim$200 kpc. Since we are interested in the azimuthal variation of the lensing signal, we check whether the boost factor has an azimuthal dependence. We present this test in Appendix \ref{app_sens}. We do not find a significant azimuthal variation in the boost correction. \\
\begin{figure}
   \includegraphics[angle=270,width=1\linewidth]{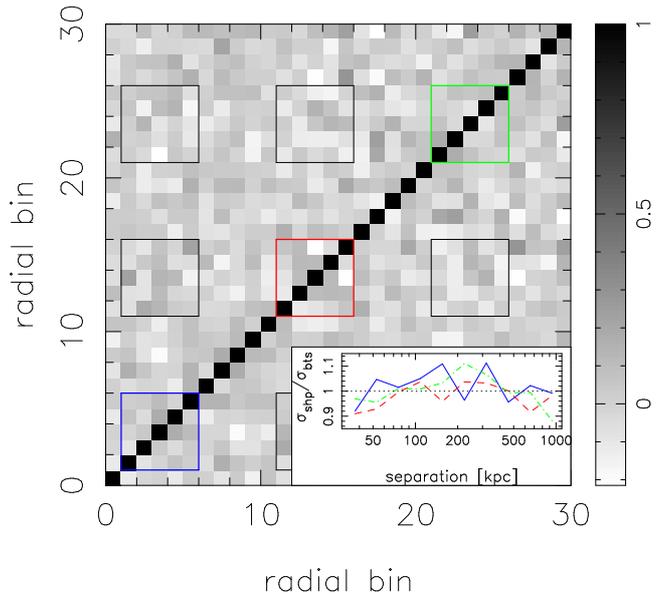}
   \caption{Correlation matrix of the lensing measurements. Bin number 1--10 corresponds to the 10 radial bins of $\Delta \Sigma$, bins 11--20 to the measurements of $\widehat{\Gamma}_{\rm t,2}$ and bins 21--30 to $\widehat{\Gamma}_{\times,2}$. The blue, red and green squares highlight the auto-correlation of the signal for the radial bins that are used in the 40 kpc $<R<$ 250 kpc fits, while the black squares show the cross-correlation between the measurements. There are no clear off-diagonals visible. The inset shows the ratio of the error on the lensing measurements determined using shape noise only and the error from bootstrapping. The blue solid line, red dashed line and green dot-dashed line correspond to  $\Delta \Sigma$, $\widehat{\Gamma}_{\rm t,2}$ and $\widehat{\Gamma}_{\times,2}$, respectively. The error estimates agree well.}
   \label{plot_cov}
\end{figure}
\begin{figure*}
   \includegraphics[angle=270,width=1\linewidth]{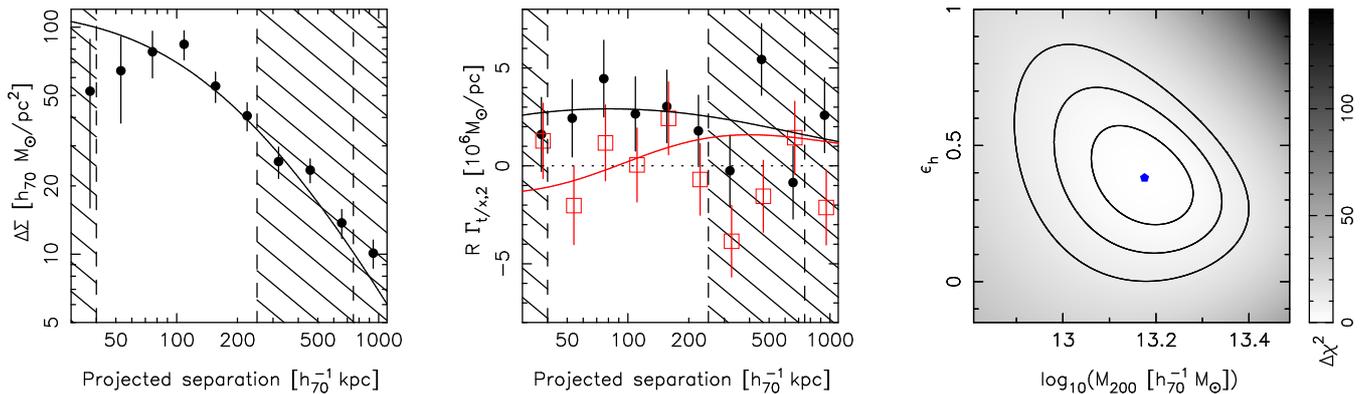}
   \caption{Isotropic weak lensing signal (left) around the brightest group members of GAMA groups with a multiplicity of $N_{\rm fof}\ge5$, and the weak lensing signal anisotropy (middle) around the major axes of the same BCGs (AX1). In the middle panel, black circles are for $\widehat{\Gamma}_{\rm t,2}$ and red squares for $\widehat{\Gamma}_{\times,2}$. The dashed vertical lines indicate the lower and upper limit of the fit range, 40 kpc and 250 kpc respectively. Solid lines are the best-fitting elliptical NFW profile. The right-hand panel shows the $\Delta \chi^2$ surface as a function of the two fit parameters, $M_{200}$ and $\epsilon_{\rm h}$. The 1-, 2- and 3-$\sigma$ contours are indicated with black lines and the best-fitting model with the blue dot.}
   \label{plot_fit}
\end{figure*}
\indent Since we are only interested in the lensing signal on $<$1 Mpc scales, we use a bootstrapping technique to determine the covariance matrix of our measurements. We measure the lensing signal on square patches of 1 deg$^2$; there are 180 non-overlapping patches in total. Next, we randomly select 180 of these patches with replacement and stack the signals to generate an approximate realization of the data. We repeat this procedure 10$^5$ times; the scatter between the bootstrap realizations approximates the measurement error. We show the correlation matrix in Fig. \ref{plot_cov}. We cannot discern significant off-diagonal terms, which is expected for a shape-noise dominated measurement. The inset of Fig. \ref{plot_cov} shows the ratio of the error computed from shape noise over the bootstrap error. The ratio is close to unity, suggesting that cosmic variance is not important on the scales that we probe. Therefore, we use the shape noise errors throughout and assume that the covariance matrix is diagonal, which enables us to use a much finer radial binning. A similar result was obtained in \citet{Viola15} for the isotropic lensing signal around GAMA groups using the KiDS-DR1/2 catalogues. Note that we use this bootstrapping method to compute the errors of the boost correction, which are much smaller than the errors on the lensing measurement and therefore ignored.

%%%%%%%%%%%%%%%%%%%%%%%%%%%%%%%%%%%%%%%%%%%%%%%%%% Results %%%%%%%%%%%%%%%%%%%%%%%%%%%%%%%%%%%%%%%%%%%%%%%%%% 

\section{Results}\label{sec_res}
We measure the lensing signal around the BCGs of GAMA groups with $N_{\rm fof}\geq5$ in the three equatorial patches, 2355 of which have a reliable shape measurement. Their average redshift is 0.22. We first adopt the major axis of the BCG as the proxy for the orientation of the halo. The lensing signal is measured in 50 logarithmically spaced bins between 30 kpc and 1100 kpc, enabling us to exactly probe the radial range of interest, and is shown in Fig. \ref{plot_fit}. For clarity, we rebin the measurements in the figures. We fit an elliptical NFW profile with a fixed mass-concentration relation to the isotropic and anisotropic part of the signal. We perform the fit in three different regimes: first, on scales between 40 kpc and 250 kpc, conservatively restricting the fit to the inner part of the halo (roughly up to $0.5 \; r_{200}$). Secondly, we use the range between 40 kpc and 750 kpc, that is approximately up to 1.5 times $r_{200}$. We choose this upper bound as 750 kpc roughly corresponds to the location where the lensing signal from neighbouring haloes becomes important \citep[see e.g.][]{VanUitert16Cl}; including their signal would complicate the interpretation. For completeness, we also fit on scales between 250 kpc and 750 kpc. We analyse the inner and outer part of the haloes separately, as \citet{Despali16} found in N-body simulations that the outer part of haloes is on average rounder and misaligned with respect to the inner part due to continuous merging events, which we can test. We note that the choice of which scale defines the inner halo is somewhat arbitrary, but we test how our results depend on it (see Fig. \ref{plot_fitrad}). \\
\indent The best-fitting elliptical NFW model for the fit on small scales is shown in Fig. \ref{plot_fit}. The right-hand panel shows the 1-, 2- and 3-$\sigma$ contours of the two fit parameters, the average halo mass and the average halo ellipticity. The marginalized constraints are \mbox{$M_{200}=1.50_{-0.24}^{+0.25} \times 10^{13} M_\odot$} and $\epsilon_{\rm h} = 0.38\pm0.12$, respectively. The halo ellipticity is therefore detected with $\gtrsim$3$\sigma$. The reduced chi-squared of the best-fitting model is $\chi_{\rm red}^2=68.3/(75-2)=0.94$ (3$\times$25 data points, 2 fit parameters), so the model provides a good fit to the data. \\
\begin{figure}
   \includegraphics[width=1\linewidth]{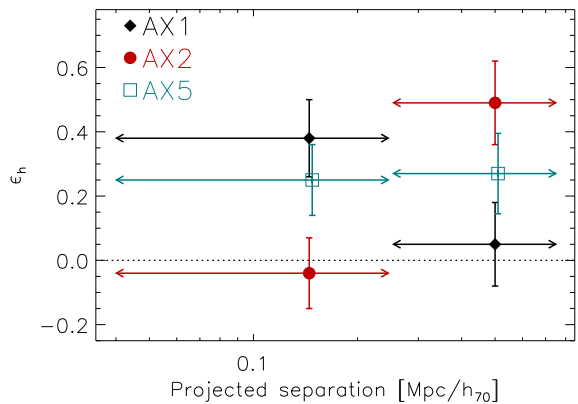}
   \caption{Constraints on the average halo ellipticity for three out of seven proxies of the orientation of the dark matter distribution that we adopt in this work. The horizontal bars with arrows show the radial range used in the fit, while the vertical error bars show the 68\% confidence interval of the halo ellipticity. The results for AX5 are slightly offset horizontally for clarity. Not shown are the halo ellipticity constraints for AX3 and AX4, which are similar to the one of AX2, and the constraints for AX6 and AX7, which are consistent with zero on all scales.}
   \label{plot_HEmain}
\end{figure}
\begin{figure}
   \includegraphics[width=1\linewidth]{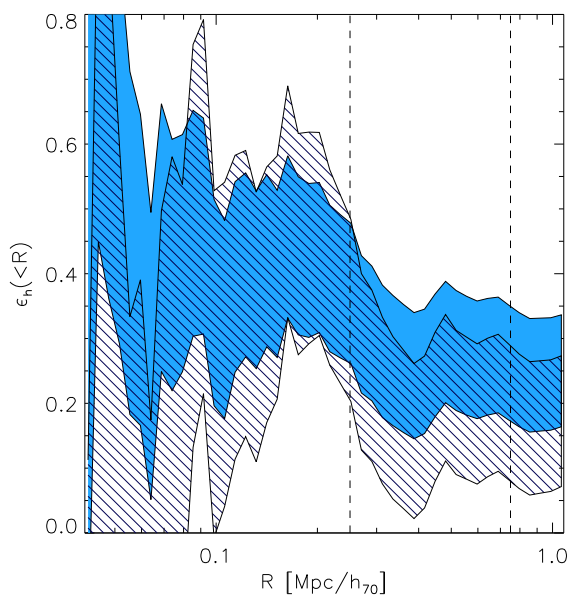}
   \caption{Constraints on the average halo ellipticity for the lensing signal anisotropy measurements of AX1, determined using all radial bins larger than 40 kpc up to the one of interest. The blue area shows the 68\% confidence regime for the fit to $\widehat{\Gamma}_{\rm t,2}$ and $\widehat{\Gamma}_{\times,2}$, while the dashed contour corresponds to the 68\% confidence regime for the fits to $\widehat{\Gamma}_{\rm t,2}+\widehat{\Gamma}_{\times,2})$. The vertical dotted lines indicate 250 kpc and 750 kpc. Note that the NFW mass was held fixed to the nominal best-fitting value. }
   \label{plot_fitrad}
\end{figure}
\indent At scales $>250$ kpc, $\widehat{\Gamma}_{\times,2}$ becomes negative, possibly indicating a fairly abrupt change in the orientation of the matter distribution, which pulls $\epsilon_{\rm h}$ down if it is included in the fit. When fitting an elliptical NFW profile at scales \mbox{250 kpc $<R<$ 750 kpc}, the average halo ellipticity is consistent with zero, as can be seen in Fig. \ref{plot_HEmain} and read off from Table \ref{tab_he}. We quantify this further by measuring \mbox{$\epsilon_{\rm h}$ ($<R$)}, that is we adopt a fixed lower bound of 40 kpc, but vary the upper bound. We keep the NFW mass fixed to its best-fitting value. The result is shown in Fig. \ref{plot_fitrad}. \mbox{$\epsilon_{\rm h}$ ($<R$)} is roughly constant up to 250 kpc, but makes a fairly sudden drop to values of $\sim$0.25 at 500 kpc, after which it remains constant towards larger scales.  \\
\indent An underprediction of $\widehat{\Gamma}_{\times,2}$ towards large radius with respect to an elliptical NFW model was also observed in the Millennium Simulation analysis in \citet{Schrabback15}, which was interpreted as being caused by shape-shear correlations \citep{Hirata04}. However, a simultaneous increase of $\widehat{\Gamma}_{\rm t,2}$ was observed there as well, for which we find no clear evidence in our data. Hence it remains unclear whether or not shape-shear correlations contribute to the measurements at these scales. \\
\indent  We have ignored the contribution from the stellar mass associated with the BCG to the lensing signal anisotropy. To verify whether this is justified, we matched our BCG sample to the publicly available single-S\'{e}rsic fit catalogue based on GAMA-DR2, version 7 \citep{Kelvin12}. The average $r$-band effective radius of our BCGs is 3.6 arcsecond, which corresponds to 12.8 kpc at the mean lens redshift. Most of the stellar mass is therefore contained within 40 kpc, the minimum scale we adopt in the fit. Even if this stellar mass component is highly elliptical, it does not affect the lensing signal anisotropy at larger scales much: the contribution of a finite mass component that is confined to small scales, drops off as $\propto \theta^{-4}$ and becomes rapidly insignificant, as can be seen from Eq. (\ref{eq_psi2simp}) and (\ref{eq_gamma_anis}).

%%%%%%%%%%%%%%%%%%%%%%%%%%%%%%%%%%%%%%%%%%%%%%%%%% Results - sensitivity %%%%%%%%%%%%%%%%%%%%%%%%%%%%%%%%%%%%%%%%%%%%%%%%%% 

\subsection{Sensitivity analysis}\label{sec_sens}
\begin{figure}
   \includegraphics[width=1\linewidth]{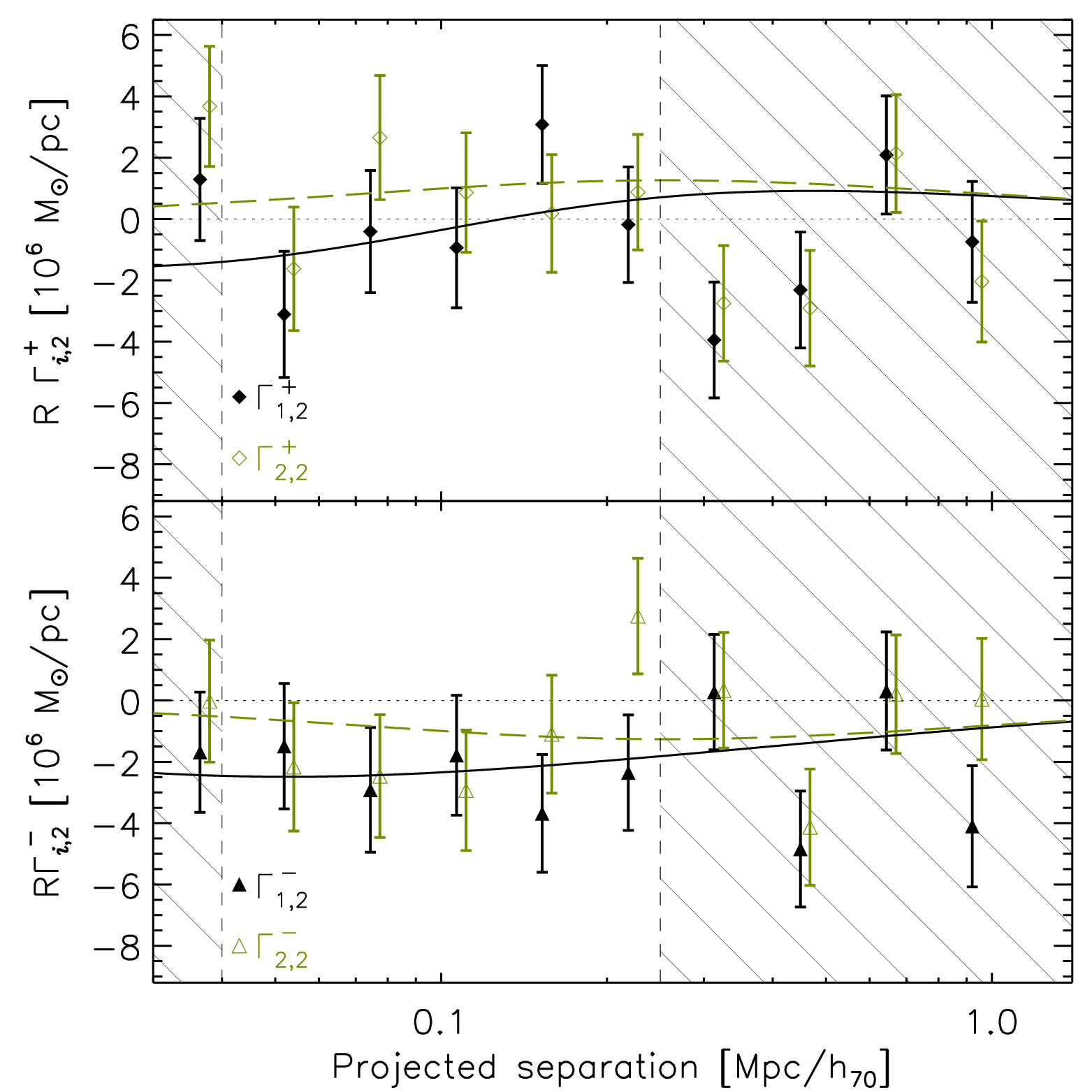}
   \caption{Anisotropic part of the lensing signal around BCGs of GAMA groups with $N_{\rm fof}\ge5$, measured using the alternative quadrupole moment estimators as defined in Eq. (\ref{eq_cl1}) to (\ref{eq_cl4}). The black solid and green dashed lines in the top (bottom) panel correspond to the best-fitting elliptical NFW profile for $\widehat{\Gamma}^{+}_{1,2}$ and $\widehat{\Gamma}^{+}_{2,2}$ ($\widehat{\Gamma}^{-}_{1,2}$ and $\widehat{\Gamma}^{-}_{2,2}$), respectively. The hatched area indicates the scales that are not used in the fit.}
   \label{plot_fitAdhi}
\end{figure}

\indent So far, we have not accounted for a potential alignment of the shapes of lenses and sources by directly fitting to Eq. (\ref{eq_Gammat2}) and Eq. (\ref{eq_Gammax2}). A spurious alignment could be caused by coherent errors in the PSF models, and by cosmic shear. Since we measured the shapes of lenses and sources with two different pipelines (with completely independent PSF models), and since the BCGs are large and bright to begin with and their shapes therefore not much affected by the PSF, we do not expect that the PSF can have a large effect. The impact of cosmic shear has been addressed with simulations in \citet{Schrabback15}. In their Fig. 6, they show the impact of cosmic shear on the lensing signal anisotropy for lenses with a stellar mass in the range $10.5 <\log_{10}(M_*)<11$ and redshifts $0.2<z<0.4$. The average impact in the range $<250$ kpc is a few per cent at most. Since our lens sample is more massive and at a lower redshift, the impact will be even smaller and can thus be safely ignored. \\
\indent However, we can test whether spurious lens-source alignments have an impact by fitting a model to \mbox{$(\widehat{\Gamma}_{\rm t,2}+\widehat{\Gamma}_{\times,2})$} instead, which removes the effect of spurious alignments altogether. As can be seen in Fig. \ref{plot_fitrad}, the resulting constraints on $\epsilon_{\rm h}$ are fully consistent over all scales. The halo ellipticity between 40 kpc and 250 kpc is \mbox{$\epsilon_{\rm h}=0.37\pm0.15$}. As a test for our model implementation, we also fit the halo ellipticity in the traditional way, as outlined in Sect. 4 of \citet{Mandelbaum06}; we obtain fully consistent results. This shows that potential lens-source alignments do not significantly bias our results. A second indication that this is the case is that lens-source alignments increasingly affect the lensing signal anisotropy towards larger projected radius \citep{Howell10,VanUitert12,Schrabback15}, which would have caused an increasing difference between the two fits when including larger scales. The small, observed increase in the difference between the results towards large scales is insignificant. \\
\indent We also measure the halo ellipticity using the alternative estimator from \citet{Clampitt16}. The measurements are shown in Fig. \ref{plot_fitAdhi}, together with the best-fitting elliptical NFW profile. We fix the halo mass to $M_{200}=1.50 \times 10^{13} M_\odot$ and only fit for the halo ellipticity. We obtain $\langle \epsilon_{\rm h} \rangle=0.38\pm0.12$. Hence the results are fully consistent. We do not expect completely identical results, because the azimuthal weighting of the lensing measurements is different and because the measurements are binned differently. The error on the halo ellipticity of this estimator is similar to the error of our fiducial method, which shows that it does not matter which method is used.   \\
\indent To test whether our results depend on the adopted mass-concentration relation, we vary the normalization of the mass-concentration relation. The results here and in the remainder of this work are again obtained by fitting Eq. (\ref{eq_Gammat2}) and Eq. (\ref{eq_Gammax2}), together with the isotropic part of the lensing signal. When we decrease the amplitude of the mass-concentration by 30\%, we obtain $\epsilon_{\rm h}=0.30_{-0.10}^{+0.11}$, while if we increase it by 30\% we get $\epsilon_{\rm h}=0.42_{-0.12}^{+0.13}$, which are both within the 1$\sigma$ errors of our fiducial constraint. By construction, the best-fitting masses also change, but since they are nuisance parameters in our fit, that is not important. \\
\indent Another potential contaminant is miscentring: although the BCG is a better estimator of the group centre compared to for example the peak of the X-ray emission or the mean position of group members \citep{George12}, some BCGs may not be the central galaxy, or some may be somewhat displaced from the bottom of the gravitational potential. Miscentring should dilute the lensing signal anisotropy. If miscentring has an azimuthal dependence (e.g. preferentially along the major axis of the dark matter distribution), which might be reasonable to expect, it becomes less straightforward to estimate the impact on halo ellipticity measurements. We address this issue by attempting to select a better centred sample. First of all, it is possible that not the brightest galaxy, but the one with the largest stellar mass, is actually the central galaxy in the group. Hence, if a satellite has a larger stellar mass than the BCG, we adopt it as the group centre instead. This happens for 19\% of the groups, but we only replace the BCG if the satellite has a reliable KSB shape (95\% have). We find that this results in $\epsilon_{\rm h}=0.38\pm0.11$, so no significant difference with our fiducial result. \\
\indent Secondly, we look at the magnitude gap between the BCG and the brightest satellite galaxy. The smaller the gap, the more likely it is that the satellite is actually the central galaxy. Hence we repeat the measurements, requiring this gap to be larger than 0.5 magnitudes. The resulting sample contains roughly half the groups. This leads to $\epsilon_{\rm h}=0.30_{-0.13}^{+0.14}$. Repeating the measurement with the other half of the sample gives $\epsilon_{\rm h}=0.51_{-0.22}^{+0.25}$, which is consistent. \\
\indent Thirdly, we also measured the signal of red BCGs only, selected using $(u-r)>1.8$. 89\% of the BCGs is red according to this criterion. The resulting halo ellipticity constraint is $\epsilon_{\rm h}=0.32_{-0.10}^{+0.11}$. \\
\indent The observed major axis of BCGs with small ellipticities may be relatively sensitive to pixel noise, which could be an additional source of scatter between the light and the dark matter, diluting the measurements. Therefore, we exclude all BCGs with $|e|<0.05$ (13\% of the sample) and repeat the measurement. The resulting halo ellipticity is $\epsilon_{\rm h}=0.38_{-0.12}^{+0.13}$, which shows that these round galaxies do not dilute the measurements. An alternative way of assessing this is by additionally weighting each lens-source pair by the lens ellipticity, following \citet{Mandelbaum06,VanUitert12,Schrabback15}. However, this leads to slightly weaker constraints, with $\epsilon_{\rm h}=0.30_{-0.13}^{+0.14}$, a downward shift of less than 1$\sigma$. This result implies that the KSB ellipticity of a galaxy is not a good indicator of the ellipticity of the projected dark matter distribution. This is not unexpected. For instance, the presence of a bulge results in a more compact weight function. In a face-on galaxy the resulting ellipticity measurement may become dominated by the bulge shape, which is not expected to be correlated with the halo ellipticity. Similarly, the ellipticity may be underestimated in a bulge-dominated edge-on galaxy. The impact of the choice of weight function will be studied in a future work (Georgiou et al., in prep.). \\
\indent We also test the sensitivity of our results on the source redshift cut. We implement a range of different cuts, compute a new $N(z)$ and boost correction for each cut, remeasure the lensing signal and refit the data. The results are presented in Appendix \ref{app_sens}. Our results are insensitive to the source redshift cut. \\
\indent Finally, we tested whether ignoring the off-diagonal terms in the covariance matrix affected the fit results. To do so, we used the full covariance matrix (Fig. \ref{plot_cov} shows the corresponding correlation matrix), inverted that and applied a correction for the bias which is introduced when a noisy covariance matrix is inverted \citep{Kaufmann67,Hartlap07}. Fitting the signal using this inverted covariance matrix, our best-fitting halo ellipticity decreased with $\sim$6\% while the 68\% confidence intervals remained unchanged. This decrease might be due to noise in the covariance matrix. In any case, it shows that replacing the full covariance matrix with a diagonal one that only includes shape noise terms does not have a significant impact on our results.

%%%%%%%%%%%%%%%%%%%%%%%%%%%%%%%%%%%%%%%%%%%%%%%%%% Results - alternative proxies %%%%%%%%%%%%%%%%%%%%%%%%%%%%%%%%%%%%%%%%%%%%%%%%%% 

\subsection{Alternative proxies for halo orientation}\label{sec_altres}

\begin{figure}
   \centering
   \includegraphics[width=1\linewidth]{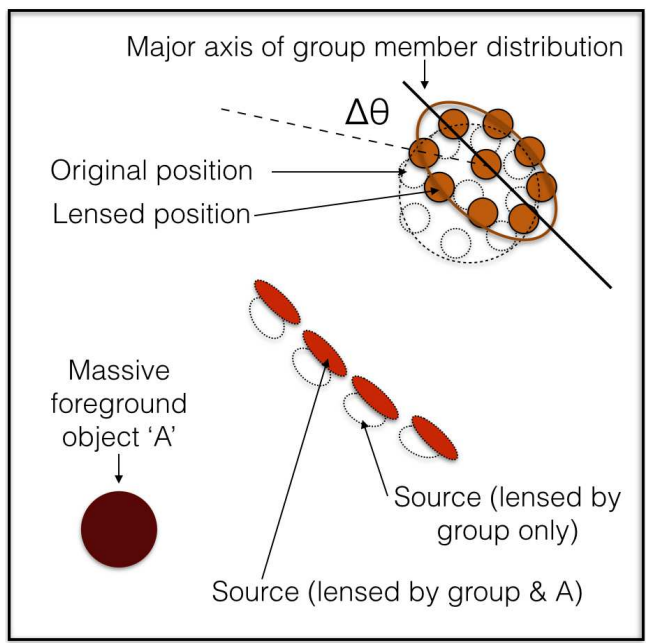}
   \caption{Cartoon of a potential bias in the halo ellipticity measurement when adopting the major axis of the satellite distribution as proxy of the orientation of the halo. Imagine a galaxy group whose member distribution is intrinsically round (dotted circles in the top right). In the presence of a massive object at a lower redshift than the group, the positions of the group members are lensed such that the group appears radially squeezed, with its observed major axis perpendicular to the separation vector. Source galaxies between the group and the massive object receive an extra shear, but now appear to reside near the minor axis of the group member distribution. This causes an apparent anti-alignment between the observed group member distribution and the inferred dark matter distribution.}
   \label{plot_ill2}
\end{figure}
The group catalogue of GAMA enables us to test various proxies for the orientation of the dark matter halo based on the distribution of the group satellites. We measure the lensing signal anisotropy using the six different proxies defined in Sect. \ref{sec_gama}. The use of proxies other than the BCG's major axis has the advantage that the measurement will not be affected by lens-source alignments due to systematics in the data such as imperfect PSF models. However, the first-order lensing effect, the actual displacement of sources rather than their shape distortion, could cause a similar bias.\\
\begin{figure*}
   \includegraphics[width=1\linewidth]{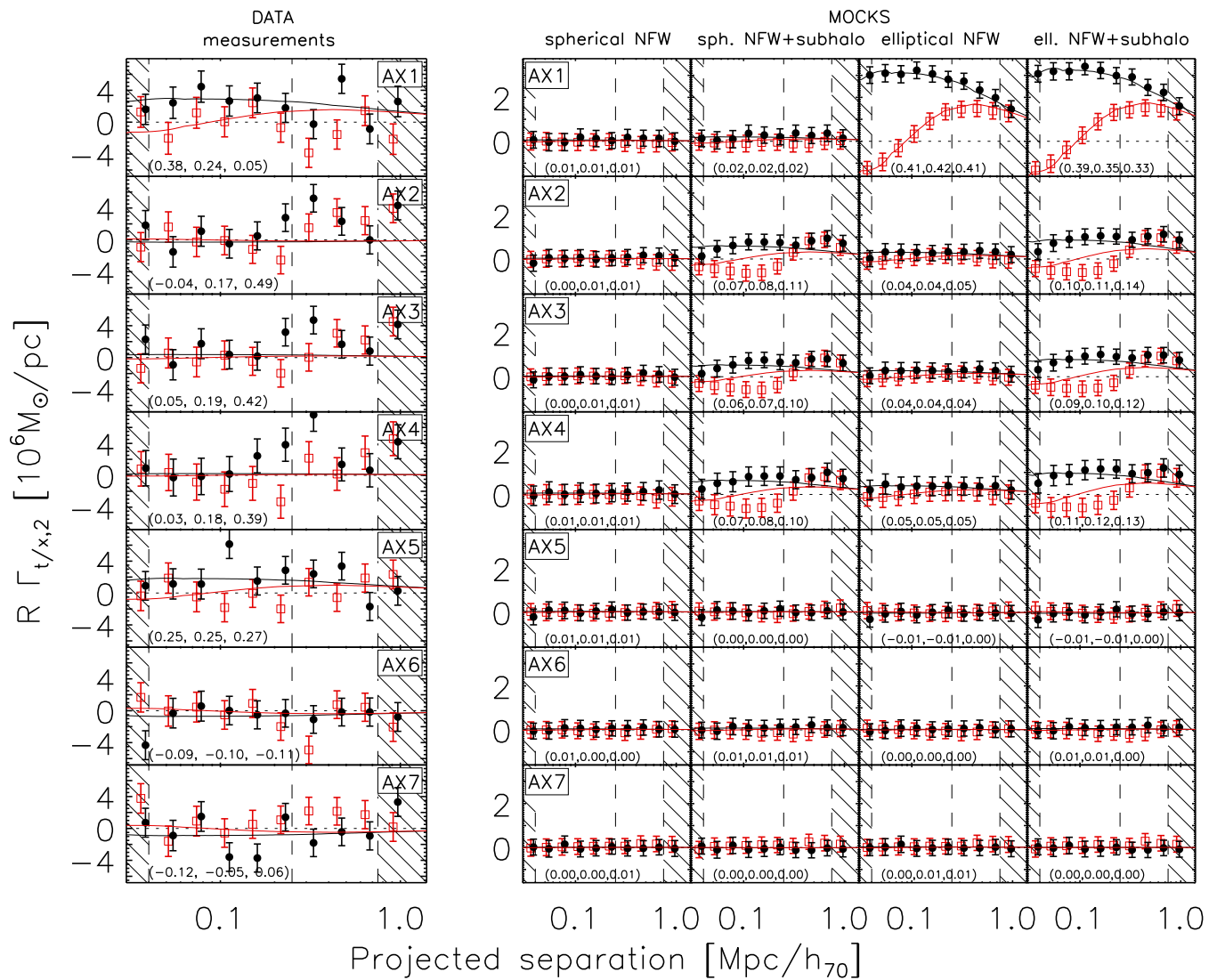}
   \caption{Anisotropic lensing signal for GAMA groups with a multiplicity of $N_{\rm fof}\ge5$ around various preferred axes. The filled black circles show $\widehat{\Gamma}_{\rm t,2}$ and the open red squares $\widehat{\Gamma}_{\times,2}$. Solid lines are the best-fitting elliptical NFW profile, fitted to the signal on scales 40 kpc $<R<$ 250 kpc. The hatched area indicates the regime that is excluded in all fits. The left-hand column shows the measurements, while the other columns show the results from mocks, as described in Sect. \ref{sec_max}. The numbers between brackets in each panel indicate the best-fitting halo ellipticities, fitted on scales of 40 kpc $<R<$ 250 kpc, 40 kpc $<R<$ 750 kpc and 250 kpc $<R<$ 750 kpc, respectively.}
   \label{plot_HEaxes}
\end{figure*}
\indent Imagine a massive, low-redshift object, next to a galaxy group at a higher redshift whose member distribution is intrinsically round. The lensing of this massive object radially squeezes the positions of the group members on the sky (just like the shape of a background galaxy is distorted due to shear), such that the observed major axis of this galaxy group appears perpendicular to the separation vector to the massive object (see Fig. \ref{plot_ill2}). Background source galaxies are lensed by the massive object and by the galaxy group, and will receive an extra shear in the region close to the apparent minor axis of the group. As a result, one expects to observe an anti-alignment between the group member distribution and the inferred dark matter distribution. Since the GAMA groups are at low redshift, there cannot be many massive structures in the foreground. Furthermore, the weight function we employ also suppresses this effect. We used simple mocks to confirm that this is a small effect and we shall ignore it. In theory, however, one could measure cosmic shear by correlating the shapes of galaxy groups as traced by their member distribution, although the signal might be dominated by intrinsic alignments (i.e. the group member distribution pointing towards neighbouring structures). \\
\indent The anisotropic part of the lensing signal is shown in the left-hand column of Fig. \ref{plot_HEaxes}. For reference, we also show the signal around the major axis of the BCG. At small scales, none of the alternative proxies give an equally clear detection of lensing signal anisotropy as the BCG's major axis. The constraints on halo ellipticity are listed in Table \ref{tab_he}. \\
\indent For AX2--AX4, the detection significance increases when we fit to 750 kpc. If we fit to large scales only, that is between 250 kpc and 750 kpc, we find a significant detection of a positive alignment between the group member distribution and the dark matter distribution. Note that the results for AX2--AX4 are highly correlated, as the scatter between their position angles is small (see Fig. \ref{plot_prefax}). The largest signal is obtained for AX2 (the distribution of all group galaxies, weighted with a Gaussian with scale-radius of 300 kpc), for which we find $\epsilon_{\rm h}=0.49\pm0.13$. Typical halo ellipticities within the virial radius are $\sim$0.3 \citep[e.g.][]{Jing02,Allgood06}, which is marginally consistent with our results, assuming that the satellites and the dark matter are well aligned. Our results have not been corrected for the dilution due to Poisson noise in estimating the group's major axes (see the discussion around Eq. \ref{eq_dil}), which would have increased the halo ellipticity by $\sim$20\%. However, there may also be factors we did not account for that could have boosted our $\epsilon_{\rm h}$, such as correlated structures along the major axis of the satellite distribution, hot gas that is aligned with the potential, and the subhaloes of satellites that are anisotropically distributed. We estimate the impact of the latter in Sect. \ref{sec_max} and will address the other factors with hydrodynamical simulations in a future work.  \\
\indent Taken at face value, our results suggest a scenario in which the satellite distribution traces the dark matter distribution at large scales, but on scales $\lesssim r_{200}$, there are effects at play that simultaneously and coherently change the distribution of the dark matter and the orientation of the BCG. Hence the local distribution of stars appears to be the best indicator of the local orientation of the dark matter, which has also been reported in hydrodynamical simulations \citep{Velliscig15}. We will discuss this further in Sect. \ref{sec_lit}. \\
\indent When we additionally weight the measurement with the ellipticity of the group member distribution, the mean ellipticity around AX2 increases to $\epsilon_{\rm h}=0.02\pm0.12$, $0.23\pm0.09$ and $0.56_{-0.14}^{+0.15}$ for the fits on scales \mbox{40 kpc $<R<$ 250 kpc}, 40 kpc $<R<$ 750 kpc and 250 kpc $<R<$ 750 kpc, respectively,  an upward shift of $\sim$0.5$\sigma$. For AX3, the upward shift of $\epsilon_{\rm h}$ is at most $\sim$0.5$\sigma$ for 250 kpc $<R<$ 750 kpc, while for AX4, the halo ellipticities do not change. \\
\indent AX5 (the vector connecting the BCG to the brightest satellite) is a compromise between AX1 and AX2 in terms of tracing the overall dark matter distribution, as it results in a tentative positive detection of $\epsilon_{\rm h}$ on all scales. For AX6 and AX7, the lensing signal is consistent with being isotropic, implying that the vectors between the BCG and the second and third brightest satellite galaxies do not trace the orientation of the underlying dark matter distribution well on any scale. The average halo ellipticity as a function of fitting range for AX2 and AX5 is shown in Fig. \ref{plot_HEmain}. \\

%%%%%%%%%%%%%%%%%%%%%%%%%%%%%%%%%%%%%%%%%%%%%%%%%% Results - alternative proxies -max like test %%%%%%%%%%%%%%%%%%%%%%%%%%%%%%%%%%%%%%%%%%%%%%%%%% 

\subsubsection{Maximum likelihood tests}\label{sec_max}

To investigate alternative explanations for the observed trends, we perform a set of maximum likelihood tests. We make mock data sets using the observed positions and redshifts of the lenses and sources. For the lenses, we use the spectroscopic redshifts and for the sources the $z_B$, but we compute the $\Sigma_{\rm crit}$ for each lens by integrating over the $N(z)$ of the sources. We start with assuming that the dark matter profile of the BCG follows a spherical NFW profile, with a mass of $M_{200}=1.50 \times 10^{13} M_\odot$, the best-fitting mass of our combined fit to AX1. We predict the shear at the location of the sources and assign those as their new ellipticities, hence we ignore shape and measurement noise. Then we analyse the mock data in the same way as the real data, that is we measure the lensing signal anisotropy around the various proxies for the orientation of the dark matter distribution. Note that this signal is practically noiseless due to the absence of shape noise in the mocks, but to ensure that the relative weighting of the different radial bins is correct when we fit a model, we assign an error that is five times smaller than the one of the corresponding radial bin in the real data. \\
\indent This first set of mock data allows us to test the impact that neighbouring groups have on the halo ellipticity. We show the resulting signals in the second column of Fig. \ref{plot_HEaxes}. Each panel also indicates the best-fitting halo ellipticity. For none of the proxies of the distribution of dark matter do we obtain a detection. Hence neighbouring groups contained in the GAMA group catalogue have no impact on the lensing signal anisotropy. \\
\indent Next, we test the impact of the subhaloes of satellite galaxies. The subhaloes cause an increase in the lensing signal along the major axis of the satellite distribution, which might bias the halo ellipticity inference of the main halo. Hence in addition to assigning an NFW profile with a mass of $M_{200}=1.50 \times 10^{13} M_\odot$ to the BCG, we assign an NFW profile with a subhalo mass of $M_{200}=7.6 \times 10^{11} M_\odot$ (5\% of the group halo mass) to each satellite in the group. This subhalo mass fraction is a factor of two larger than what has been observed for satellites in GAMA \citep{Sifon15} and therefore exaggerates the effect that subhaloes might have. We reassign source ellipticities and redo the measurements using the various proxies, as shown in the third column of Fig. \ref{plot_HEaxes}. For AX2--AX4, we find a small but non-zero lensing signal anisotropy. The $\widehat{\Gamma}_{\rm t,2}$ signal looks similar to an elliptical NFW profile, but the sign of the $\widehat{\Gamma}_{\times,2}$ term is flipped. The resulting best fitting halo ellipticities are small but positive. Hence the subhaloes may cause a bias of the order of a few per cent, but the effect is much smaller than the signal measured around AX2--AX4.  \\
\indent In our third set of mocks, we assign an elliptical NFW profile to the BCG of the same mass as before with $ \epsilon_{\rm h}=0.4$, perfectly aligned with the major axis of the BCG. Subhalo masses are set to zero again. In this case, we obtain a clear signal for AX1, as shown in the fourth column of Fig. \ref{plot_HEaxes}. The scatter between this proxy and the others is so large, that most of the lensing signal anisotropy is washed out when we measure the lensing signal anisotropy around the other proxies. Finally, we create a mock where we additionally assign spherical subhaloes to the satellites with masses of $M_{200}=7.6 \times 10^{11} M_\odot$. Again, as Fig.  \ref{plot_HEaxes} shows, the most clear detection comes from adopting AX1, as in the data. Hence these mocks show that neighbouring groups and subhaloes of satellites are unlikely to affect the halo ellipticity estimates, in the scenario where the major axis of the BCG closely traces the distribution of dark matter. We also experimented with a subhalo mass fraction that increases with distance to the BCG, but we could not find a prescription that mimicked the observed trends.

%%%%%%%%%%%%%%%%%%%%%%%%%%%%%%%%%%%%%%%%%%%%%%%%%% Results - literature comparison %%%%%%%%%%%%%%%%%%%%%%%%%%%%%%%%%%%%%%%%%%%%%%%%%% 

\subsection{Literature comparison and discussion}\label{sec_lit}

\indent This work is the first weak lensing study of halo shapes that specifically targets spectroscopically identified galaxy groups. A number of studies exist in the literature in which the lensing signal anisotropy of objects with comparable masses are presented. We shall compare our results to those. It is important to keep in mind, however, that the object selection between these studies is different, which makes a detailed comparison impossible.  \\
\indent \citet{Clampitt16} measured the lensing signal anisotropy around a sample of $70\,000$ luminous red galaxies (LRGs), as well as around a sample of $2\,700$ clusters from the redMaPPer cluster catalogue \citep{Rykoff14}, using shape measurement catalogues from the SDSS. The LRGs have a similar average mass to our lens sample, while the cluster sample is ten times more massive. The adopted proxy for the orientation of the dark matter is the major axis of the lens galaxy (our AX1). The signal is fitted on scales of 0.05 $<R<$ 4 $h^{-1}$Mpc, hence extending to larger radii than in our work. For the LRGs, they report $\langle \epsilon_{\rm h} \rangle=0.24\pm0.06$, while for the redMaPPer clusters, they find $\langle \epsilon_{\rm h} \rangle=0.21$ with 3$\sigma$ significance. To compare this with our results, we measure the lensing signal anisotropy around AX1 using their original estimators (which corresponds to their halo ellipticity definition of $(1-q^2)/(1+q^2)$) and fit the signal up to 1 Mpc. We obtain $\langle \epsilon_{\rm h} \rangle=0.52\pm0.19$, which is $\sim$1.5$\sigma$ higher than their results for LRGs.\\
\indent In \citet{Evans09}, a subsample of relatively isolated clusters of the maxBCG cluster sample \citep{Koester07} was studied. They stack the lensing signal around the major axis of the satellite distribution, defined as in our Eq. (\ref{eq_esat}), but without a radial weight function. This proxy is most similar to our AX3. In their analysis, they mask the region inside $R<500$ $h^{-1}$kpc from the cluster centre, because of concerns about the robustness of the lensing signal measured at those scales. They fit an elliptical NFW profile and find an axis ratio of $q=0.48_{-0.09}^{+0.14}$, which corresponds to $\epsilon_{\rm h} / \epsilon_{\rm mem}=1.37_{-0.26}^{+0.35}$, where $\epsilon_{\rm mem}$ is the average ellipticity of the galaxy member distribution, defined as $(1-q)/(1+q)$. If we fit our AX3 signal with the same lower limit but with an upper limit of 1 Mpc, we find $\epsilon_{\rm h}=0.74_{-0.30}^{+0.35}$, which corresponds to  $\epsilon_{\rm h} / \epsilon_{\rm mem} =2.12_{-0.87}^{+1.00}$, which is consistent but noisier, because their fits extend to larger scales, and because their lens sample is larger and on average about an order of magnitude more massive than ours. Also, we should remind the reader that the halo ellipticities become increasingly biased towards larger values, and  $\epsilon_{\rm h}>1$ is formally ruled out. \\
\indent \citet{Oguri12} constrained the average halo ellipticity of 28 massive strong-lensing clusters from the Sloan Giant Arc Survey \citep{Hennawi08} and the SDSS, using a weak lensing analysis of follow-up observations from Subaru/Suprime-cam. They used the position angle from strong lensing as the major axis estimate to stack the weak lensing signal. Converting their results to our definition of halo ellipticity, they found $\epsilon_{\rm h}=0.31\pm0.05$, determined on scales $<3$ Mpc/$h$, which agrees well with our constraints on scales $<1$ Mpc/$h_{70}$. The halo ellipticity was also determined on smaller scales, but this did not significantly affect the best-fitting value. \\
\indent \citet{Schrabback15} constrained halo shapes of samples of galaxies from CFHTLenS \citep{Heymans12}. The most comparable lens sample is comprised of elliptical galaxies in the range $M_*>10^{11}$, for which they report $\epsilon_{\rm h}/\epsilon_{\rm g}=-0.09\pm0.38$. This implies that the haloes are much rounder than the galaxies, or that a substantial misalignment exists between these red galaxies and their dark matter haloes. \citet{VanUitert12} analysed a sample of red galaxies in the second Red-sequence Cluster Survey \citep{Gilbank11,VanUitert11} and also reported no detection. Finally, \citet{Mandelbaum06} measured the lensing signal anisotropy around galaxies in the SDSS. For their brightest sample of red lenses, they report $\epsilon_{\rm h}/e_{\rm g}=1.7\pm0.7$ by fitting the signal on scales 40 $h^{-1}$kpc to 300 $h^{-1}$kpc, hence a weak detection of a positive alignment. Note that none of these galaxy lens samples described in the paragraph consisted of central galaxies only. \\
\indent To summarize the above: the observational evidence for non-zero halo ellipticities of galaxy-scale haloes is still elusive, possibly because of a larger level of misalignment between the lenses and their haloes and possibly also because of satellites in the lens samples diluting the anisotropic lensing signal. Several papers including ours, however, have now reported significant non-zero halo ellipticities for group- and cluster-scale haloes. These detections imply that the dark matter distribution is well aligned with the lens light and/or the distribution of satellites in this mass range. \\
\indent A change in the orientation and ellipticity of dark matter haloes as a function of scale has been reported in \citet{Despali16}, who used cosmological N-body simulations to compare the ellipticities of dark matter haloes for a range of overdensity criteria. They found that the outer part of relaxed haloes is rounder than the inner part due to continuous merging events, and that the typical misalignment angle between the orientation of the inner part of the halo and the halo within the virial radius, is $\sim$20 deg. We cannot directly compare our result to these predictions, as we measure the signal at separate radial scales, while they reported the misalignment angle as a function of overdensity, which mixes scales. However, it might provide an explanation for our findings. \\
\indent Hydrodynamical simulations form another interesting set of results to compare to, particularly because the misalignment distribution of  galaxies and dark matter appears to depend on baryonic feedback prescriptions \citep{Tenneti16}, which implies that halo ellipticity measurements, in combination with predictions from dark-matter-only simulations, could be used to constrain baryonic physics. A number of works have investigated the relation of the alignment between galaxies and their dark matter hosts. \citet{Tenneti15} studied the alignment between galaxies and dark matter in the MassiveBlack-II simulations \citep{Khandai15}. They report that the probability distribution of the miscentring angles between the stars and the dark matter within $R_{200}$, the equivalent of our AX1 measurement, peaks at 0$^{\circ}$ and drops to almost zero at $\sim$30$^{\circ}$ for massive haloes. This corresponds to a Gaussian misalignment distribution with a standard deviation of $10^{\circ}-15^{\circ}$, which dilutes the halo ellipticity by (Eq. \ref{eq_dil}) $6-13$\%. Furthermore, they report an increased alignment between the stellar component and the dark matter distribution towards smaller radii, in qualitative agreement with our AX1 results. \\
\indent \citet{Velliscig15} study the alignment between stars and dark matter in the EAGLE \citep{Schaye15} and cosmo-OWLS \citep{LeBrun14} hydrodynamical simulations. They find that the misalignment between the stellar distribution inside a given radius, relative to the overall orientation of the dark matter halo, increases towards smaller radii, where the two components probe different scales. However, when both the stellar distribution and the dark matter are determined as a function of enclosed radius, they find a very tight alignment between the two components. This result agrees well with our findings. On small scales, the BCG dominates the stellar budget, and hence we find it to be a good indicator for the orientation of dark matter halo inside 250 kpc. Towards larger scales, the satellites dominate the stellar budget, hence there, the satellite distribution forms a better tracer of the orientation of the dark matter. This also agrees with the results from \citet{Shao16}, who use the EAGLE simulation to study the distribution of satellites in dark matter haloes. They report that satellites are better aligned with the entire halo than with its inner part, and that satellites therefore preferentially trace the outer halo. \\
\indent To optimize the weak lensing signal anisotropy measurements, it might therefore be ideal to use as proxy the orientation of the stellar distribution as a function of enclosed radius, using the light from the BCG and the other group members. \\

%%%%%%%%%%%%%%%%%%%%%%%%%%%%%%%%%%%%%%%%%%%%%%%%%% Results - Ellipticity of the satellite distribution %%%%%%%%%%%%%%%%%%%%%%%%%%%%%%%%%%%%%%%%%%%%%%%%%% 

\subsection{Ellipticity of the satellite distribution around BCGs}\label{sec_es}

\begin{figure}
   \includegraphics[angle=270,width=1\linewidth]{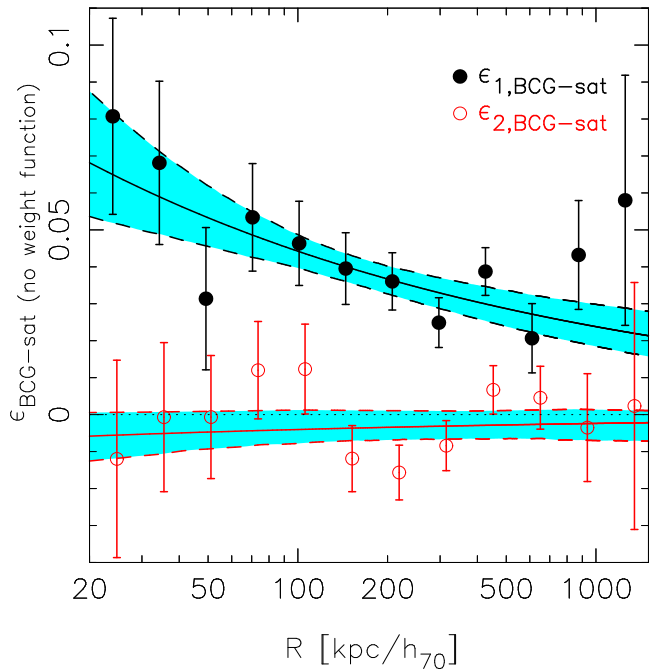}
   \caption{Ellipticity of the satellite distribution around the major axis of the BCGs. Solid lines show the best-fitting power laws, while the cyan regions show the 68\% model uncertainty. There is weak evidence that $e_{\rm 1,BCG-sat}$ decreases with radius. }
   \label{plot_esat}
\end{figure}
\indent An alternative way to gain insight in the relative orientation of BCGs, satellites and dark matter, is by measuring the ellipticity of the satellite distribution with respect to the major axis of the BCGs. We select all satellites in a concentric ring around the BCGs, rotate the system such that the BCG's major axis is aligned with the horizontal axis, and determine the ellipticity using Eq. (\ref{eq_esat}), but without adopting a radial weight function. If satellites preferentially reside near the BCG's major axis, $\epsilon_{\rm1,BCG-sat}$ should be positive while $\epsilon_{\rm2,BCG-sat}$ should be consistent with zero. Note that we added `BCG' to the subscript to make clear that these ellipticities are measured with respect to the BCG's major axis. The results are shown in Fig. \ref{plot_esat}. To determine the errors, we created bootstrap realisations of the data by randomly drawing BCGs (with their satellites) from the total sample with replacement. The scatter between the bootstrap realisations determines the error. We also implemented an alternative method to estimate errors, that is by randomizing the position angles of the BCGs. Both approaches led to very similar error bars and covariance matrices. \\
\indent $\epsilon_{\rm1,BCG-sat}$ appears to become smaller towards larger radii. To quantify this, we fit a power law of the form $\epsilon_{\rm BCG-sat}=A_{\rm sat}\times(R/100 [{\rm kpc}/h_{70}])^{\alpha_{\rm sat}}$. We used the full covariance matrix in the fit which we determined from the bootstrap realisations. For $\epsilon_{\rm 1,BCG-sat}$, we obtain \mbox{$A_{\rm sat}=0.044\pm0.005$} and $\alpha_{\rm sat}=-0.27\pm0.11$, hence providing weak evidence that the mean ellipticity of the satellite distribution becomes smaller at larger separations from the BCG. Such a trend could either imply that the distribution changes, or that the misalignment increases at larger scales. $\epsilon_{\rm 2,BCG-sat}$ is consistent with zero as expected. For comparison, the average total ellipticity of the brightness distribution of the BCGs from KSB is $\sim$0.16. It would be interesting to repeat this measurement on hydrodynamical simulations, to investigate to what extent this measurement is sensitive to baryonic physics. 

%%%%%%%%%%%%%%%%%%%%%%%%%%%%%%%%%%%%%%%%%%%%%%%%%% Conclusions %%%%%%%%%%%%%%%%%%%%%%%%%%%%%%%%%%%%%%%%%%%%%%%%%% 

\section{Conclusions}\label{sec_conc}

We measured the average isotropic and anisotropic part of the weak gravitational lensing signal around more than $2\,600$ GAMA galaxy groups with five or more members, using the shape measurement catalogues of KiDS. The anisotropic part of the lensing signal was measured separately adopting seven different proxies of the a priori unknown orientation of the dark matter halo, including the major axis of the light from the BCG, as well as various proxies based on the distribution of satellites in the group. \\
\indent On small scales ($<250$ kpc), we detect a lensing signal anisotropy around the BCG, but not around the other proxies, which implies that the BCG's major axis is the optimal proxy of the orientation of the dark matter halo on small scales. To relate that to an halo ellipticity, we have to adopt a model density profile. We derive new and simple expressions to compute the quadrupole moments of the lensing signal for any elliptical surface mass density profile. Under the assumption that the alignment between light and dark matter is perfect and the dark matter follows an elliptical NFW profile, we derive an average dark matter halo ellipticity of $\epsilon_{\rm h}=0.38\pm0.12$, in fair agreement with predictions of $\Lambda$CDM based dark-matter-only simulations \citep{Jing02,Allgood06}, which predict typical halo ellipticities of $\sim$0.3. A narrow misalignment distribution between stars and dark matter for objects of this mass is supported by hydrodynamical simulations \citep{Tenneti15}. We verified our results with alternative model implementations. \\
\indent On scales $>250$ kpc, no systematic alignment of the lensing signal with the orientation of the BCG is discovered. When we stack the lensing signal around the major axis of the satellite distribution, however, we find that the lensing signal on those large scales is anisotropic and consistent with an elliptical NFW profile. The signal is most significant for our AX2 proxy, which is the major axis of the group member distribution, determined using a Gaussian weight function with a projected scale radius of 300 kpc. The resulting halo ellipticity on those scales is $\epsilon_{\rm h} = 0.49\pm0.13$. \\
\indent We investigated various possible explanations for the observed lensing signal anisotropy. We created mocks based on the data to test the impact of neighbouring groups and satellite subhaloes, and found that their impact is much smaller than what is observed. Also, our data does not show signs of a spurious lens-source alignment, either caused by inaccurate PSF removal in the shape measurement process or caused by cosmic shear. Such an alignment, if present, would have affected the two components of the lensing signal anisotropy measurements with opposite sign and with an increasing amplitude as a function of radius, which we do not observe in our data. \\
\indent Our results, therefore, point to a scenario in which satellites trace the large-scale dark matter distribution, but on small scales, physical effects are at play which change the distribution of dark matter and simultaneously affect the orientation of the BCG, keeping their relative alignment intact. This scenario agrees well with results from hydrodynamical simulations, where it was found that the stellar distribution enclosed within a certain radius forms a good estimator for the orientation of the dark matter within the same radius \citep{Velliscig15}. \\
\indent Although not the main focus of this work, we have also reported results on the distribution of satellites in galaxy groups. We found that the satellites preferentially reside near the major axis of the BCG, in line with previous results from the literature. Also, we measured the ellipticity of the satellite distribution with respect to the major axes of the BCGs, and found weak evidence that it decreases towards larger separations. \\
\indent Our detection of a lensing signal anisotropy for a relatively small lens sample that covers only a small fraction of the sky, highlights the potential of such measurements with future data sets. With surveys such as LSST \citep{LSST09} and Euclid \citep{Laureijs11}, we can expect weak-lensing anisotropy measurements with percent level precision. Combined with predictions from dark-matter-only simulations, the lensing signal anisotropy becomes a powerful tool to constrain the misalignment distribution, which itself can be used to constrain different scenarios of baryonic feedback during structure formation \citep[e.g.][]{Tenneti16}. Additionally, the radial dependence of halo ellipticity will enable stringent tests of modified gravity models and will enable competitive constraints on the cross-section of the dark matter particle within a self-interacting dark matter framework. 

\paragraph*{Acknowledgements}
We thank the referee for his/her comments that improved the draft. EvU acknowledges support from an STFC Ernest Rutherford Research Grant, grant reference ST/L00285X/1. HHo acknowledges support from the European Research Council under FP7 grant number 279396. BJ acknowledges support by an STFC Ernest Rutherford Fellowship, grant reference ST/J004421/1. This work is supported by the Deutsche Forschungsgemeinschaft in the framework of the TR33 `The Dark Universe'. AC acknowledges support from the European Research Council under the FP7 grant number 240185. CH acknowledges support from the European Research Council under grant number 647112. HHi is supported by an Emmy Noether grant (No. Hi 1495/2-1) of the Deutsche Forschungsgemeinschaft. KK acknowledges support by the Alexander von Humboldt Foundation. RN acknowledges support from the German Federal Ministry for Economic Affairs and Energy (BMWi) provided via DLR under project no. 50QE1103. MV acknowledges support from the European Research Council under FP7 grant number 279396 and the Netherlands Organisation for Scientific Research (NWO) through grants 614.001.103. This work is based on data products from observations made with ESO Telescopes at the La Silla Paranal Observatory under programme IDs 177.A-3016, 177.A-3017 and 177.A-3018. GAMA is a joint European-Australasian project based around a spectroscopic campaign using the Anglo-Australian Telescope. The GAMA input catalogue is based on data taken from the Sloan Digital Sky Survey and the UKIRT Infrared Deep Sky Survey. Complementary imaging of the GAMA regions is being obtained by a number of independent survey programs including GALEX MIS, VST KiDS, VISTA VIKING, WISE, Herschel-ATLAS, GMRT and ASKAP providing UV to radio coverage. GAMA is funded by the STFC (UK), the ARC (Australia), the AAO, and the participating institutions. The GAMA website is http://www.gama-survey.org/. \\
\indent {\it Author Contributions}: All authors contributed to the development and writing of this paper. The authorship list reflects the lead author (EvU) followed by two alphabetical groups. The first alphabetical group includes those who are key contributors to both the scientific analysis and the data products. The second group covers those who have either made a significant contribution to the data products, or to the scientific analysis.
\bibliographystyle{mnras}

%%%%%%%%%%%%%%%%%%%%%%%%%%%%%%%%%%%%%%%%%%%%%%%%%% APPENDIX %%%%%%%%%%%%%%%%%%%%%%%%%%%%%%%%%%%%%%%%%%%%%%%%%% 

\begin{appendix}

\section{Source redshift selection}\label{app_sens}

\begin{figure*}
   \includegraphics[width=1\linewidth]{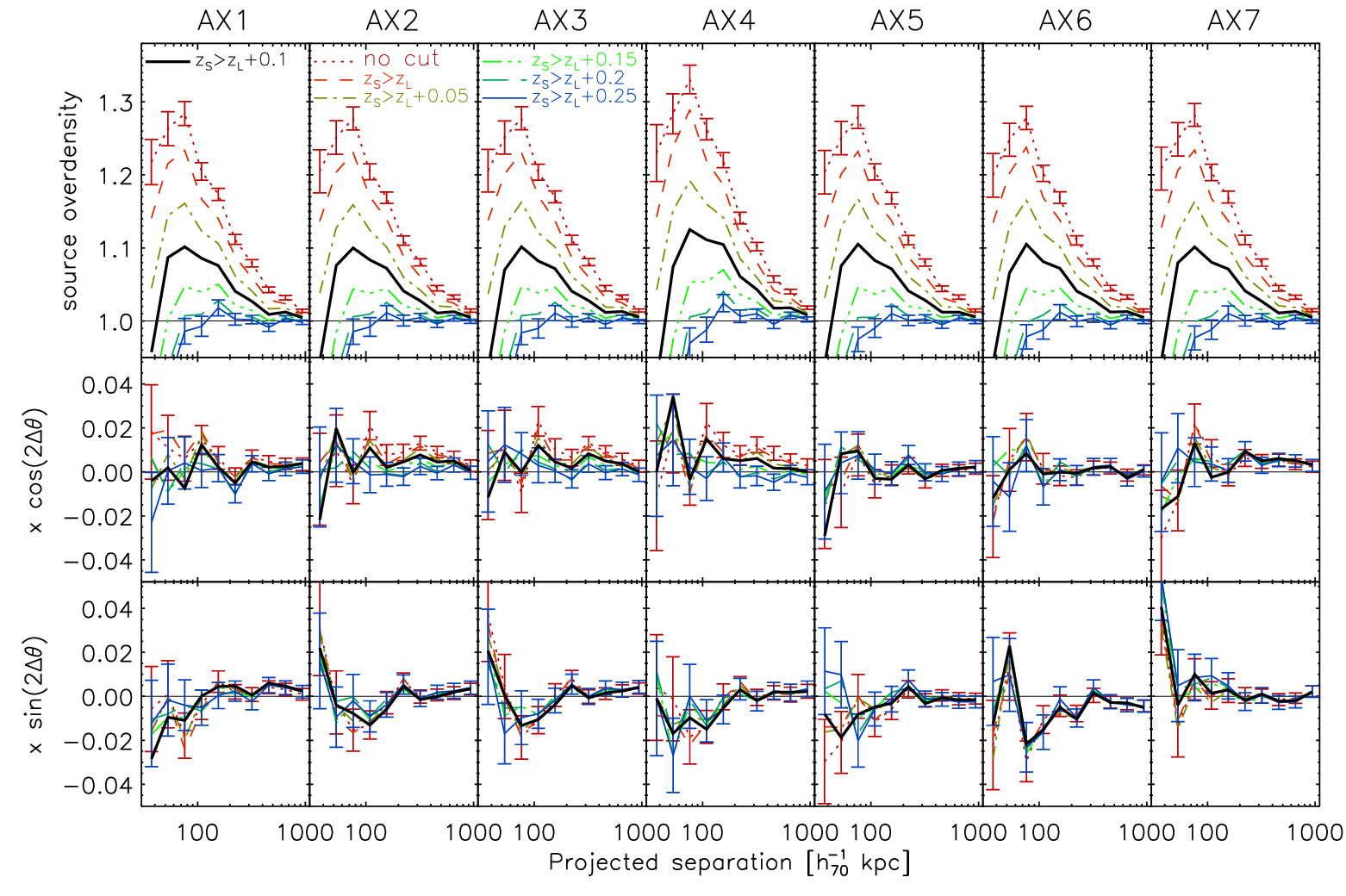}
   \caption{The overdensity of KiDS source galaxies around GAMA galaxy groups with $N_{\rm fof}\ge5$. The various lines correspond to different redshift cuts applied to the source sample. Even for a fairly conservative cut of $z_B$(source)$>z_L+0.1$, we find a residual contamination of group members in the source sample of up to 10\% at 75 kpc for our AX1 sample. }
   \label{plot_od}
\end{figure*}
\begin{figure}
   \includegraphics[width=1\linewidth]{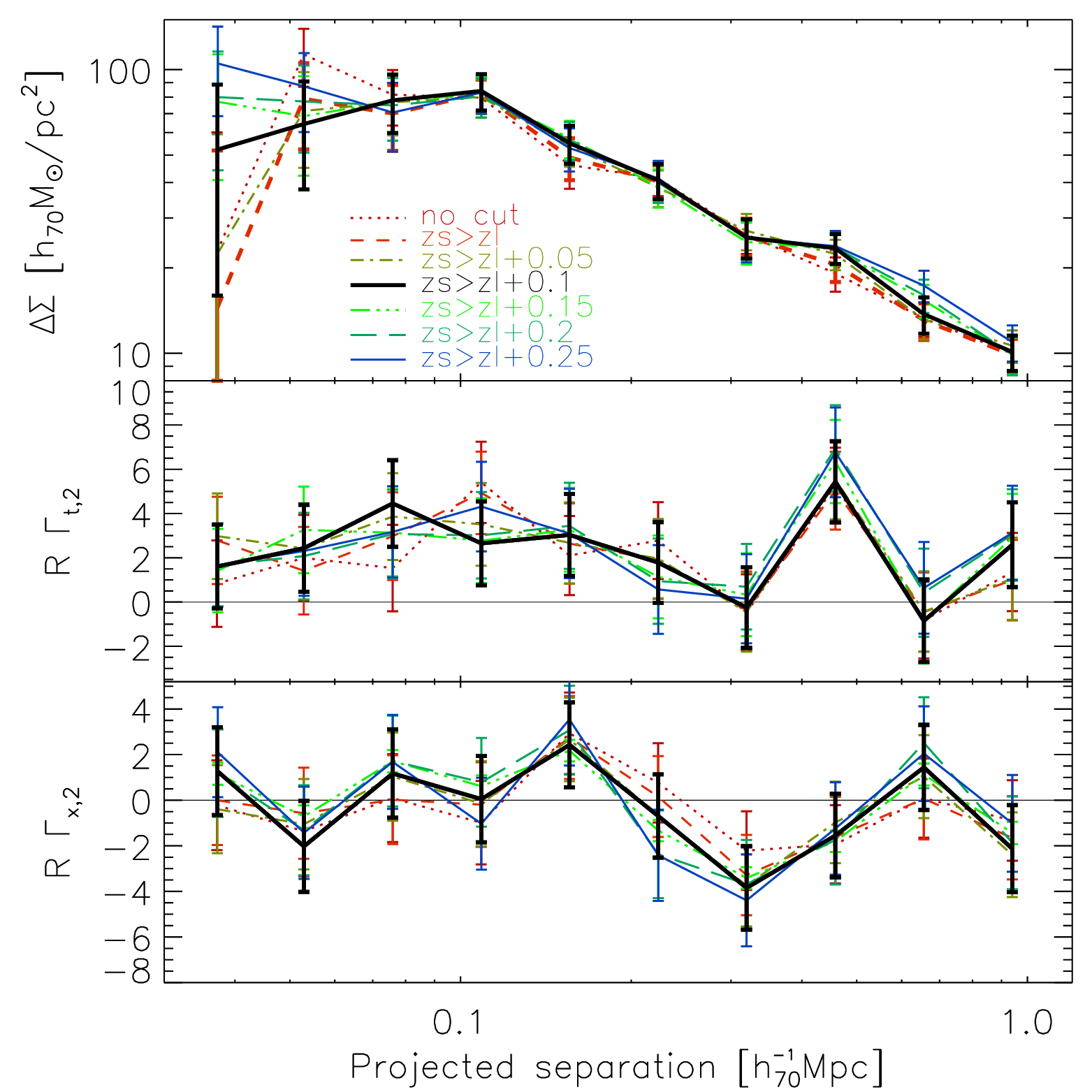}
   \caption{Isotropic and anisotropic part of the lensing signal of GAMA groups with $N_{\rm fof}\geq5$ around AX1, for different source redshift cuts. Units of the vertical axis of the lower two panels are $10^6 M_\odot$/pc. For each cut, we account for the different lensing efficiencies and contamination corrections. The lensing signal anisotropy is robustly measured against changes in the source redshift cut.}
   \label{plot_robtest}
\end{figure}

In our measurements, we apply a source redshift cut to minimize the contamination of galaxies that are physically associated with the lenses. The fiducial cut we apply is $z_B$(source)$>z_L+0.1$, with $z_B$ the Bayesian point estimate of the photometric redshift \citep{Benitez00}. The probability distribution of individual source redshifts is rather broad, however, hence we can expect some level of contamination, even though our lenses are at much lower redshifts than our sources. Since these physically associated sources are not lensed, they bias the lensing signal low if unaccounted for. \\
\indent We measure this contamination by determining the relative density of source galaxies around the lens with respect to the background density. The results are shown in the top row of Fig. \ref{plot_od}. The different columns in Fig. \ref{plot_od} show the results for the different proxies of the orientation of the dark matter distribution. A signal larger than unity shows that the source sample contains galaxies that are associated with the lens. For AX1, we find that after our fiducial redshift cut, the contamination can be as high as 10\% at scales of $\sim$75 kpc. The overdensity quickly drops towards larger scales. The contamination can be further suppressed by applying more aggressive cuts, but this is not ideal as real source galaxies are removed as well, which decreases the lensing signal-to-noise ratio. \\
\indent Contamination biases the lensing signal low by a factor that is equal to the overdensity; hence we can correct the lensing signal by multiplying it with the factors shown in this figure. Hence contamination is not an issue for this study, as long as it is isotropic. If all the physically associated galaxies would preferentially reside near the major axis of the lens, the lensing signal in that direction would be more diluted, which would bias the lensing signal anisotropy low. To determine whether that is the case, we also determine the anisotropic part of the contamination correction, shown in the second and third row of Fig. \ref{plot_od}. Note that the measurements are fairly correlated, hence the trends that are apparent in some of the panels of the lower two rows are less significant than they appear. The azimuthal dependence of the contamination is less than a percent at all scales, which is much smaller the errors of the lensing measurements of the corresponding bins. Hence we do not take it into account in our analysis. \\
\indent As a robustness test, we repeat the halo anisotropy measurements around the major axis of the BCG (AX1) for the different source redshift cuts. For each cut, we recompute the lensing efficiencies and the contamination correction. The isotropic and anisotropic part of the lensing signal are shown in Fig. \ref{plot_robtest}. The lensing signal anisotropy does not depend on the particular value of the redshift cut, from which we conclude that it is robustly measured. We present the halo ellipticity constraints in Table \ref{tab_check}, fitted on scales \mbox{40 kpc $<R<$ 250 kpc}. The different redshift cuts do not lead to significant changes in the results. It is interesting to note that the errors on the average halo ellipticity in Table \ref{tab_check} do not depend much on the source redshift cut. Applying a less stringent redshift cut increases the number of sources and hence decreases the errors on the lensing measurement, but this is largely canceled by a larger boost factor, which increases both the lensing signal and its error.

\begin{table}
  \caption{Halo ellipticity constraints for various redshift cuts, obtained from simultaneous fits to the isotropic and anisotropic part of the lensing signal of GAMA groups with $N_{\rm fof}\geq5$ around AX1 on scales 40 kpc $<R<$ 250 kpc.}   
  \begin{tabular}{c c c} 
  \hline
  Redshift cut & $\epsilon_{\rm h}$ & $\epsilon_{\rm h}$ ($\widehat{\Gamma}_{\rm t,2}+\widehat{\Gamma}_{\times,2}$)\\
  \hline\hline
\\
 no cut & $0.38\pm0.12$ & $0.38\pm0.15$ \\
 $z_S>z_L$  & $0.42_{-0.12}^{+0.13}$ & $0.42_{-0.15}^{+0.16}$ \\
 $z_S>z_L+0.05$  & $0.41_{-0.12}^{+0.13}$ & $0.40_{-0.15}^{+0.16}$ \\
 $z_S>z_L+0.1$  & $0.38\pm0.12$ & $0.37\pm0.15$ \\
 $z_S>z_L+0.15$  & $0.36_{-0.11}^{+0.12}$ & $0.40\pm0.15$ \\
 $z_S>z_L+0.2$  & $0.34_{-0.12}^{+0.13}$ & $0.39_{-0.15}^{+0.16}$ \\
 $z_S>z_L+0.25$  & $0.34_{-0.12}^{+0.13}$ & $0.36\pm0.16$ \\
\\
  \hline
  \end{tabular}
  \label{tab_check}
\end{table} 

\end{appendix}

\clearpage

\end{document}